\documentclass[aps,amsmath,amssymb,prb,10pt,twocolumn,showpacs,letterpaper]{revtex4-1}
\usepackage{graphicx,txfonts,color,tcolorbox}
\usepackage{bm}
\usepackage{mathrsfs}
\usepackage{comment}
\usepackage{mathtools}
\usepackage{hyperref}
\usepackage{epstopdf}
\usepackage{letltxmacro}
\usepackage{appendix}

\newcommand{\al}{\alpha}
\newcommand{\be}{\beta}
\newcommand{\g}{\gamma}
\newcommand{\de}{\delta}

\newcommand{\z}{\zeta}

\newcommand{\thi}{\theta}

\newcommand{\la}{\lambda}

\newcommand{\p}{\pi}

\newcommand{\s}{\sigma}

\newcommand{\W}{\Omega}
\newcommand{\De}{\Delta}

\newcommand{\ta}{\tau}

\newcommand{\pd}{\partial}

\newcommand{\round}[1]{\left( #1 \right)}
\renewcommand{\square}[1]{\left[ #1 \right]}
\newcommand{\curly}[1]{\left\{#1\right\}}

\newcommand{\beq}{\begin{equation}}
\newcommand{\eeq}{\end{equation}}
\newcommand{\Beq}{\begin{eqnarray}}
\newcommand{\Eeq}{\end{eqnarray}}
\newcommand{\bml}{\begin{multline}}

\newcommand{\bea}{\begin{align}}
\newcommand{\ena}{\end{align}}
\newcommand{\bsp}{\begin{split}}
\newcommand{\esp}{\end{split}}

\newcommand{\nn}{\nonumber}

\newcommand{\bp}{{\boldsymbol p}}

\newcommand{\bj}{{\boldsymbol j}}

\newcommand{\bk}{{\boldsymbol k}}

\DeclareMathOperator{\sgn}{sgn}

\newcommand{\hH}{\hat{H}}

\newcommand{\hS}{\hat{S}}

\newcommand{\hbS}{\hat{\boldsymbol{S}}}

\newcommand{\bde}{\boldsymbol{\de}}
\newcommand{\hpsi}{\hat{\psi}}

\newcommand{\hc}{\hat{c}}
\newcommand{\ve}{\varepsilon}

\newcommand{\hJ}{\hat{\mathrm{J}}}

\newcommand{\hb}{\hat{b}}

\newcommand{\hI}{\hat{I}}

\newcommand{\hT}{T_K}

\newcommand{\hrho}{\hat{\rho}}

\newcommand{\hvf}{\hat{\varphi}}
\newcommand{\hy}{\hat{\phi}}
\newcommand{\htvf}{\hat{\tilde{\varphi}}}
\newcommand{\df}[1]{{\color{red}@#1}}

\begin{document}
\title{Noise in tunneling spin current across coupled quantum spin chains}
\author{Joshua Aftergood}
\author{So Takei}
\affiliation{Department of Physics, Queens College of the City University of New York, Queens, NY 11367, USA}
\affiliation{Physics Doctoral Program, The Graduate Center of the City University of New York, New York, NY 10016, USA}
\date{\today}

\begin{abstract}
We theoretically study the spin current and its dc noise generated between two spin-$1/2$ spin chains weakly coupled at a single site in the presence of an over-population of spin excitations and a temperature elevation in one subsystem relative to the other, and compare the corresponding transport quantities across two weakly coupled magnetic insulators hosting magnons. In the spin chain scenario, we find that applying a temperature bias exclusively leads to a vanishing spin current and a concomitant divergence in the spin Fano factor, defined as the spin current noise-to-signal ratio. This divergence is shown to have an exact analogy to the physics of electron scattering between fractional quantum Hall edge states and not to arise in the magnon scenario. We also reveal a suppression in the spin current noise that exclusively arises in the spin chain scenario due to the fermion nature of the spin-1/2 operators. We discuss how the spin Fano factor may be extracted experimentally via the inverse spin Hall effect used extensively in spintronics.
\end{abstract}
\maketitle

\section{Introduction} 
\label{intro}
The quantification of spin-dependent charge current  noise\cite{tserkovPRB01,mishchenkoPRB03,lamacraftPRB04,belzigPRB04,forosPRL05,zareyanEPL05,chudnovskiyPRL08,meairPRB11,arakawaPRL15} as well as pure spin current noise\cite{wangPRB04,sauretPRL04,kamraPRB14} in mesoscopic conductors has garnered much attention over the past two decades, demonstrating the importance of spin effects on charge transport. 
In contrast, the study of pure spin current noise in insulating spin systems (i.e., quantum magnets) has received only limited attention.~\cite{kamraPRL16,kamraCM17,matsuoCM17} This focal imbalance, however, may soon resolve with the recent pioneering developments in spintronics, where experimentalists are now capable of generating and detecting pure spin currents in insulating magnets using purely electrical signals.~\cite{kajiwaraNAT10,cornelissenNATP15,goennenweinAPL15,liNATC16} In these experiments, two strongly spin-orbit coupled metals are affixed to two opposite ends of a magnetic insulator [e.g., yttrium iron garnet (YIG)], charge current is passed through one metal generating spin current in the magnet via the spin Hall effect (SHE), and charge current is detected in the second metal generated by the inverse SHE. These advancements open doors to the fascinating possibility to quantify spin propagation through quantum magnets via electrical measurements, and render the theoretical investigation of pure spin current noise in these systems timely.

\begin{figure}[t]
\includegraphics[width=\linewidth]{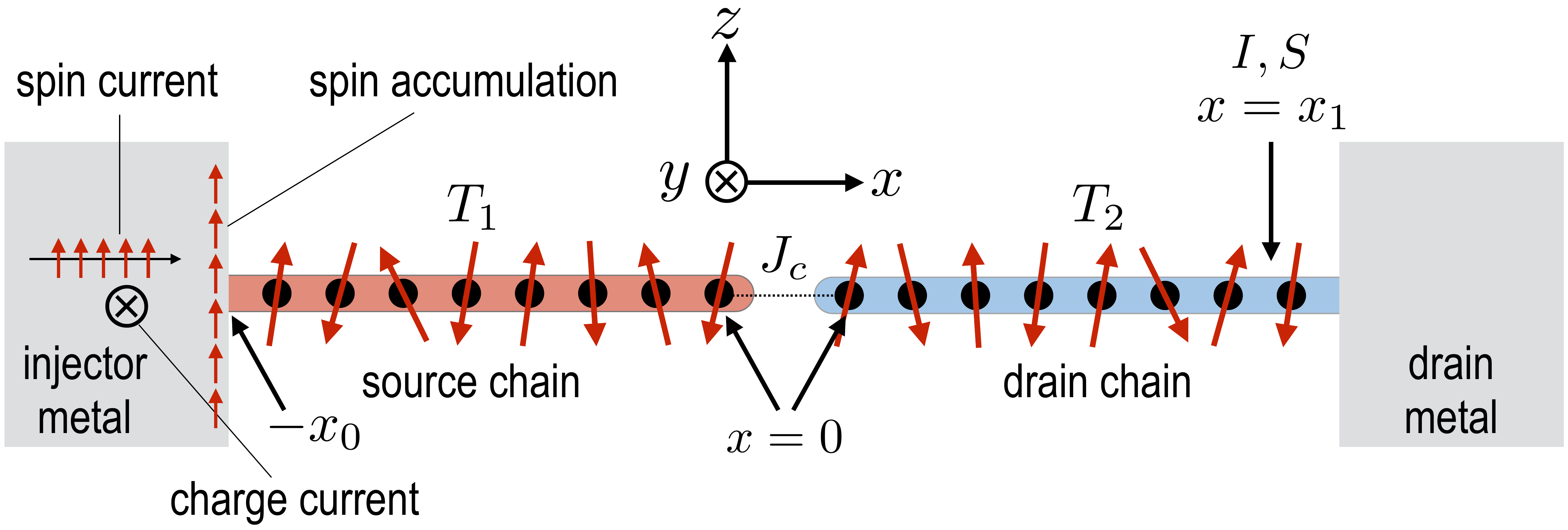}
\caption{(Color online) Two semi-infinite quantum spin chains, one labeled the source and the other the drain, are coupled at their finite ends. Strongly spin-orbit coupled metals act as injector and drain for spin current. The source chain is held at temperature $T_1$ and the drain chain at $T_2$. The spin injection from the injector metal via the spin Hall effect leads to a nonequilibrium accumulation of spin in the source chain. Depicted positions are continuum variables.}
\label{fig1}
\end{figure}

A natural setup to study spin current and noise in quantum magnets involves two quantum magnets weakly coupled via the exchange interaction (see, e.g., Fig.~\ref{fig1}). In the presence of a bias, the exchange coupling allows spin-1 excitations to stochastically tunnel from one system to the other, generating a noisy spin current in the latter. In this context, the physics of spin injection into a quantum magnet should depend on the spin quantum number $s$ of the localized spins. If a spin-1 excitation is injected into an $s=1/2$ quantum magnet, a second spin-1 excitation cannot be injected at the same site, generating a partial blockade (or {\em Pauli blockade}) during spin injection associated with the fermionic nature of the spin-1/2 operators.~\cite{jordanZP28} Pauli blockade should be absent in large-$s$ quantum magnets, where an approximate theoretical description of the injection process in terms of tunneling bosonic quasiparticles (i.e., magnons) is appropriate. This crossover from boson-like to fermion-like spin injection physics as $s$ approaches the quantum limit should have an effect on the tunneling spin current and noise and have direct experimental consequences on spin transport.

In this work, we compare spin transport across weakly-coupled $s=1/2$ quantum magnets to that across weakly-coupled large-$s$ magnetic insulators and evaluate quantities that differentiate between the two: the spin current, its dc noise and the {\em spin Fano factor} defined as the noise-to-signal ratio of the spin current. Specifically, for a quantum magnet we consider the $s=1/2$ antiferromagnetic quantum spin chain (QSC) due to its amenability to rigorous theoretical analysis~\cite{klumperBOOK04,*mikeskaBOOK04} and relevance to real materials.~\cite{hammarPRB99,*stonePRL03,*lancasterPRB06,*kuehnePRB09,*kuehnePRB11,*klanjsekP14,*hirobeNATP17} 
We consider spin currents generated by an over-population of spin excitations in one quantum magnet (i.e., subsystem) relative to the other while simultaneously applying a temperature difference between the two subsystems. We find a strikingly different behavior in the spin Fano factor between the QSC and the large-$s$ (magnon) cases. Unlike the magnon case, a vanishing spin current and concomitant diverging spin Fano factor is found only in the QSC scenario even in the presence of a large temperature bias, provided no over-population of spin excitations in one subsystem relative to other exists. We show 
that this finding is in exact analogy with the results obtained in the physics of local electron scattering between edge states of two fractional quantum Hall liquids.~\cite{heiblumPSS06,kaneBOOK07} We also compare the noise generated by nonequilibrium injection of spin into one subsystem to that generated by injecting spins into both subsystems, and find that over-populating both subsystems actually reduces the noise in the QSC scenario but increases the noise in the magnon case. We attribute this suppression in the noise to Pauli blockade physics associated with the fermionic nature of spin-1/2 operators in the QSC scenario. We finally discuss how the spin Fano factor may be experimentally obtained via the inverse SHE used in spintronics.

This paper is organized as follows. Sec.~\ref{model} introduces our model for the quantum spin chain, wherein we describe the Luttinger model and further reveal a striking similarity between our problem and a similar situation in fractional quantum Hall physics. In Sec.~\ref{keld}, we utilize a general nonequilibrium scheme based on the Keldysh formalism to derive spin transport properties in our system of interest, and contrast it to the same calculation in a geometrically equivalent large-$s$ magnonic system. We discuss our results in Sec.~\ref{disc}, and present a possible method for experimental verification in Sec.~\ref{exp}. Finally, we conclude in Sec.~\ref{conc} and propose potential avenues for future exploration.

\section{Model} 
\label{model}
Our system of interest is comprised of two $s=1/2$ antiferromagnet spin chains exchange-coupled end-to-end, one labeled the source chain ($\nu = 1$) and the other the drain chain ($\nu = 2$), with each additionally coupled to a metal with strong spin-orbit coupling (e.g., Pt, Ta, etc.) at the other ends (see Fig.~\ref{fig1}). To allow for possible spin transfer across the spin chains, we consider elevating the temperature of one QSC relative to the other (i.e., {\em thermal bias}) and/or injecting a $z$ polarized spin current (hereafter simply referred to as spin current) into the source QSC (i.e., {\em chemical bias}). The spin current injection can be facilitated, e.g., by SHE at the upstream end of the chain, where the injector metal is coupled and driven by a charge current. Due to spin-orbit coupling in the metal, a charge current flowing in the $y$ direction can give rise to a spin current flowing in the $x$ direction with spin polarized in the $z$ direction and produce a ($z$-polarized) spin accumulation at the interface (as shown in Fig.~\ref{fig1}).~\cite{maekawaBOOK12,sinovaRMP15} Interfacial exchange interaction then allows spin angular momentum to be transferred from the metal's electron spins to the spin moments in the QSC, effectively leading to an injection of spin current into the QSC.~\cite{kajiwaraNAT10,cornelissenNATP15,goennenweinAPL15,liNATC16,brataasNATM12,haneyPRB13,tserkovnyakPRB14} The injected spin current tunnels across the QSCs and is eventually ejected into the drain metal, where the spin current converts into a transverse charge current via the inverse SHE and can therefore be detected electrically. Our focus will be on the spin current flowing at $x=x_1$ in the drain chain and its dc noise (see Fig.~\ref{fig1}), which should be electrically detectable using the drain metal. 

\subsection{Luttinger model for the spin chains} 
\label{ll}
We consider two identical $s=1/2$ semi-infinite $xxz$ antiferromagnet chains that are weakly exchange-coupled at their finite ends, i.e., at site $j=0$ (or $x=0$), as shown in Fig.~\ref{fig1}. The total Hamiltonian for the QSCs can then be written as $\hH=\sum_{\nu=1,2}\hH_\nu+\hH_b+\hH_c\equiv \hH_0+\hH_b+\hH_c$, where the Hamiltonians of the two QSCs read
\beq
\label{hnu}
\hH_\nu=J\sum_{j}\curly{\hS^x_{\nu,j}\hS^x_{\nu,j+1}+\hS^y_{\nu,j}\hS^y_{\nu,j+1}+\De\hS^z_{\nu,j}\hS^z_{\nu,j+1}}\ ,
\eeq
with $J,\De>0$, and the coupling Hamiltonian reads
\beq
\hH_{c}=J^\perp_c\curly{\hS^{x}_{1,0}\hS^{x}_{2,0}+\hS^{y}_{1,0}\hS^{y}_{2,0}}+J^z_c\hS^{z}_{1,0}\hS^{z}_{2,0}\ ,
\label{hc}
\eeq
where we assume $|J^\perp_c|,\ |J^z_c|\ll J$ and we are allowing for an $xxz$ anisotropy in the exchange coupling. The bias Hamiltonian $\hH_b$ will be specified later. 

We assume $\De<1$ such that the QSCs are in the $xy$ phase with unaxial symmetry and possess a gapless excitation spectrum.~\cite{mikeskaBOOK04} 
In this regime and in the long-wavelength limit, Eq.~\eqref{hnu} is well-described by a Luttinger liquid Hamiltonian~\cite{gogolinBOOK98,mirandaBJP03,giamarchiBOOK04}
\beq
\begin{multlined}
\label{llh}
\hH_0=\frac{\hbar u}{4\pi K}\int_{-\infty}^0 dx\curly{[\pd_x\htvf_{1,R}(x)]^2+[\pd_x\htvf_{1,L}(x)]^2}\\ 
+\frac{\hbar u}{4\pi K}\int^{\infty}_0 dx\curly{[\pd_x\htvf_{2,R}(x)]^2+[\pd_x\htvf_{2,L}(x)]^2}\ ,
\end{multlined}
\eeq
where $u$ is the speed of the chiral boson fields, $K$ is the so-called Luttinger parameter, and the chiral boson fields obey $[\htvf_{\nu,R}(x),\htvf_{\nu',R}(x')]=-[\htvf_{\nu,L}(x),\htvf_{\nu',L}(x')]=i\p K\de_{\nu\nu'}\sgn(x-x')$.~\cite{gogolinBOOK98} 
Arriving at Eq.~\eqref{llh} requires that we drop RG-irrelevant operators (e.g., band-curvature and backscattering terms) and that we constrain our inquiry to the Gaussian regime. In order to remain within the Gaussian regime we find $u$ and $K$ perturbatively by taking $\De\ll 1$. The result is that $u = v_F/K$ and $K\simeq 1 - 2\De/\pi$ for the low energy sector. However, a Bethe ansatz approach shows that the Gaussian model in fact holds for the entire critical domain $|\De|<1$ provided we identify $u=\p v_F\sqrt{1-\De^2}/2\cos^{-1}(\De)$ and $K=[2-(2/\p)\cos^{-1}(\De)]^{-1}$.~\cite{johnsonPRA73} Therefore, given this exact solution, we assume throughout that $\De\lesssim1$ and so take $u\simeq\p v_F/2$ and $K\gtrsim 1/2$, i.e., close to the Heisenberg limit with $\De\lesssim1$.


Spin injection at the upstream end of the source QSC should generate a spin accumulation in the QSC, which, in the long-time (steady-state) limit, can be modeled as a (uniform) spin chemical potential $\mu$ that extends over its entire length, causing the spins to precess about the $z$ axis and driving current across the coupling. 
The Hamiltonian describing the chemical bias may then be written as
\beq
\label{hb}
\hH_{b}=\frac{\mu}{2\p}\int_{-\infty}^{0}dx\ \partial_x(\htvf_{1,R}+\htvf_{1,L})\ .
\eeq

\subsection{Exact mapping to the problem of stochastic electron tunneling between two fractional quantum Hall edge channels}
\label{fqh}
In order to treat the semi-infinite chains we must account for the finite boundary. We begin this process by first introducing new scaled chiral fields $\hvf_{\nu,R/L}=\htvf_{\nu,R/L}/\sqrt{K}$ in Eq.~\eqref{llh}, allowing us to map this system onto an effectively non-interacting (i.e., $K = 1$ or free fermion) Luttinger liquid governed by the Hamiltonian
\beq
\begin{multlined}
\label{sfh}
\hH_0=\frac{\hbar u}{4\pi}\int_{-\infty}^0 dx\curly{[\pd_x\hvf_{1,R}(x)]^2+[\pd_x\hvf_{1,L}(x)]^2}\\ 
+\frac{\hbar u}{4\pi}\int^{\infty}_0 dx\curly{[\pd_x\hvf_{2,R}(x)]^2+[\pd_x\hvf_{2,L}(x)]^2}\ ,
\end{multlined}
\eeq
where now the scaled chiral fields obey $[\hvf_{\nu,R}(x),\hvf_{\nu',R}(x')]=-[\hvf_{\nu,L}(x),\hvf_{\nu',L}(x')]=i\p \de_{\nu\nu'}\sgn(x-x')$. Then the semi-infinite boundary conditions at $x = 0$ requires that $\hvf_{\nu,R}(0) = -\hvf_{\nu,L}(0)$, which further enforces that the string operator $\cos{(\cdots)} \rightarrow 1$ at the end.~\footnote{For a full explication of a finite edge in a Luttinger liquid, see section $10.1$ of Ref.~\onlinecite{giamarchiBOOK04}.} We can extend the $x=0$ result to include all space and time by noting that right-movers are a function of $x-ut$ only and left-movers are a function of $x+ut$ only. Thus we have 
\beq
\label{bc}
\hvf_{\nu,R}(-x,t)=-\hvf_{\nu,L}(x,t)\ .
\eeq 

We now impose Eq.~\eqref{bc} explicitly on Eq.~\eqref{sfh} and reinstate the unscaled fields $\htvf_{\nu,R}(x)$ such that
\beq
\label{hnus2}
\hH_0=\frac{\hbar u}{4\p K}\sum_{\nu=1,2}\int_{-\infty}^\infty dx\ [\pd_x\htvf_{\nu,R}(x)]^2\ .
\eeq
We note that the remaining right chiral fields now reside on an infinite domain and obeys the commutation relation $[\htvf_{\nu,R}(x),\htvf_{\nu',R}(x')]=i\p K\de_{\nu\nu'}\sgn(x-x')$; they can be expanded in terms of canonical boson operators as
\beq
\label{fnur}
\htvf_{\nu,R}(x)=-i\sqrt{\frac{2 \pi K}{L}}\sum_{k>0} \frac{e^{-\eta k /2}}{\sqrt{k}} \bigg\{\hb_{\nu,k}e^{ikx}-\hb^\dag_{\nu,k}e^{-ikx}\Big\}\ ,
\eeq
where $\eta$ is a UV cutoff and $L$ is the chain length (eventually taken to infinity). The boson operator $\hb_{\nu,k}$ diagonalizes Eq.~\eqref{sfh} as $\hH_\nu=\sum_{k>0}\ve_k\hb^\dag_{\nu,k}\hb_{\nu,k}$ with $\varepsilon_k=\hbar uk$. 

Explicitly implementing Eq.~\eqref{bc} on the spin chemical bias term Eq.~\eqref{hb}, we obtain
\beq
\label{hb2}
\hH_b=\frac{\mu}{2\p}\ \square{\int_{-\infty}^{0}dx\ \pd_x\htvf_{1,R}(x)+\int^{\infty}_{0}dx\ \pd_x\htvf_{1,R}(x)}\ .
\eeq
Finally, applying Eq.~\eqref{bc} on the bosonized spin operators we find
\beq
\label{bso}
\hS^-_{\nu,0}=\frac{\sqrt{a}\round{\g_{\nu,R}+\g_{\nu,L}}}{\sqrt{2 \pi \eta}}e^{\frac{i}{K}\htvf_{\nu,R}(0)}= (\hS^+_{\nu,0})^\dag\ ,
\eeq
where $a$ is the lattice constant for the spin chain, $\g_{\nu}$ are Majorana fields that obey the anti-commutation relation $\{\g_{\mu},\g_{\nu}\} = 2\de_{\mu \nu}$. Using Eq.~\eqref{bso}, the coupling Hamiltonian Eq.~\eqref{hc} can now be re-expressed as
\beq
\label{hc2}
\hH_c=\xi_\perp e^{\frac{i}{K}[\htvf_{1,R}(0)-\htvf_{2,R}(0)]}+h.c.\ ,
\eeq
where $\xi_\perp\equiv (J^\perp_ca/4\p\eta)(\g_{1,R}+\g_{1,L})(\g_{2,R}+\g_{2,L})$. Equation~\eqref{hc2} should in principle contain the $z$ component of the exchange coupling that gives rise to a term proportional to $J^z_c[\pd_x\htvf_{1,R}(0)][\pd_x\htvf_{2,R}(0)]$. However, a leading-order RG analysis gives that the scaling dimension for the coupling $\xi^\perp$ is $1-1/K$ while that for $J^z_c$ is $-1$. Since we assume $K>1/2$, the latter term is less RG-relevant than the terms appearing in Eq.~\eqref{hc2} so in the long-wavelength low-energy limit, the inter-chain coupling should be dominated by the transverse components of the exchange coupling presented in Eq.~\eqref{hc2}.

\begin{figure}[t]
\includegraphics[width=0.7\linewidth]{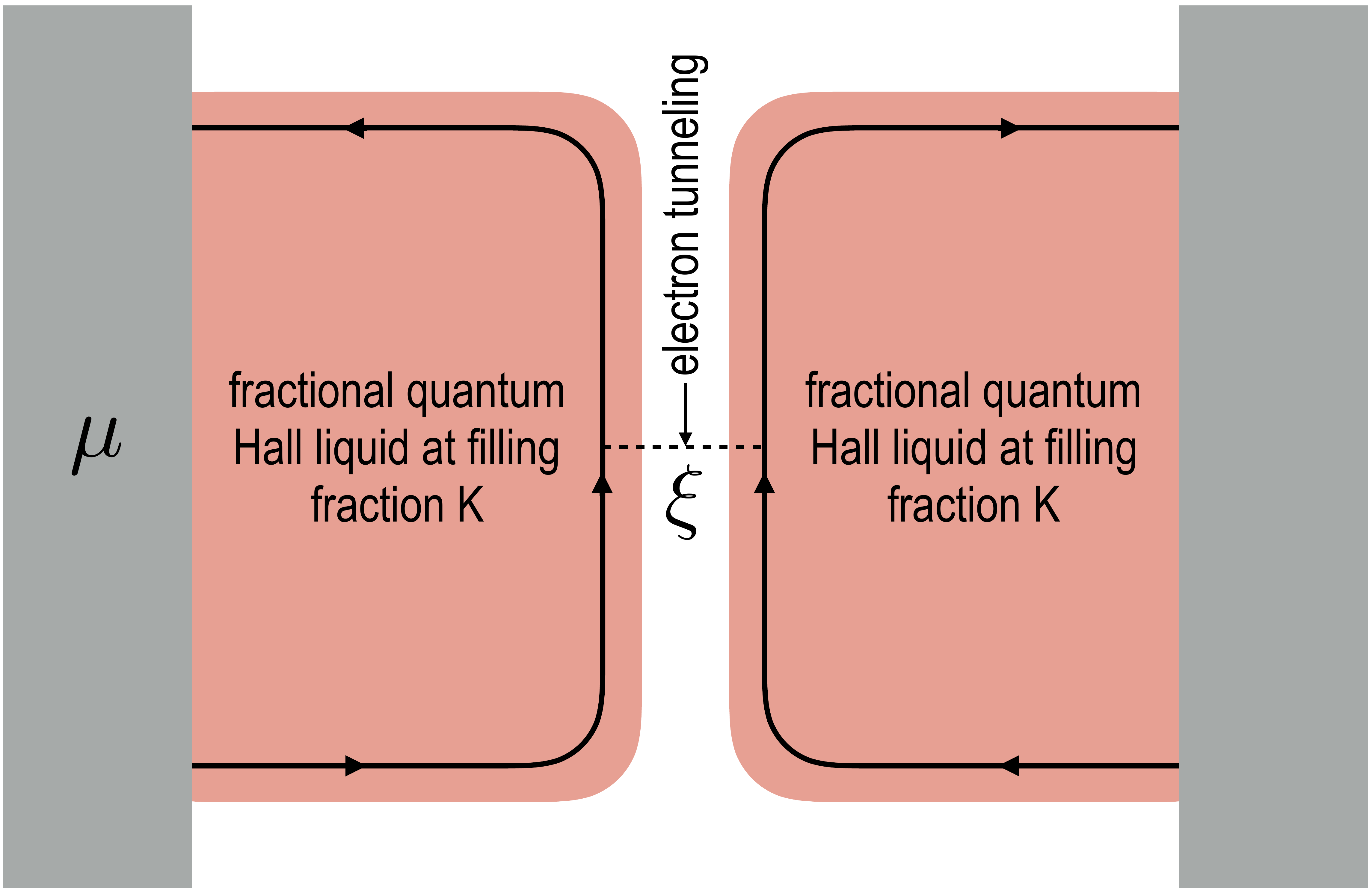}
\caption{(Color online) The problem of tunneling spin-1 excitations between two semi-infinite QSCs exactly maps to the problem of stochastic electron tunneling between the edge states of two Laughlin fractional quantum Hall liquids at filling fraction $K$. The role of the spin chemical bias in the QSC problem is played by the electrical voltage $\mu= eV$ applied to the edge channel on the left in the quantum Hall case.}
\label{fig2}
\end{figure}

We note that Eqs.~\eqref{hnus2}, \eqref{hb2} and \eqref{hc2} exactly correspond to a theoretical model describing two-terminal charge transport between the edge channels of two ``Laughlin" fractional quantum Hall liquids at filling fraction $K$ (see Fig.~\ref{fig2}).~\cite{kaneBOOK07} In the fractional quantum Hall scenario, Eq.~\eqref{hc2} describes a stochastic tunneling of (charge $-e$) electrons (with amplitude $\xi_\perp$) between the two edge states, and the role of the spin chemical bias $\mu$ is played by the (electrical) voltage bias $\mu=eV$ applied between the two edge channels. Equation~\eqref{hnus2} resembles the edge state Hamiltonian for the Laughlin fractional quantum Hall liquids, but we note that for the actual Laughlin states, $K$ is directly determined by the topological property of the bulk quantum Hall state and is constrained to inverse odd integers.~\cite{kaneBOOK07}

\section{Spin current and dc noise} 
\label{keld}
We now focus on the spin current flowing in the drain chain and its dc noise at the spatial point $x=x_1$ just to the left of the drain metal (see Fig.~\ref{fig1}). These transport quantities may be measured electrically in the drain metal. The spin current can be obtained by evaluating the Keldysh expectation value of the operator for spin current [after implementing the boundary condition \eqref{bc}]
\beq
\label{cur}
\begin{aligned}
\hI(x_1,t)=\lim_{x\rightarrow x_1}\frac{\hbar u}{2\p}\sum_{\rho=\pm}\rho\pd_x\htvf_{2,R}(\rho x,t)\equiv\sum_{\rho=\pm}\hI_\rho(x_1,t)\ .\\
\end{aligned}
\eeq
In the infinite past, we assume the source (drain) QSC to be in a thermal state with respect to $\hH_1+\hH_{b}$ ($\hH_2$) at temperature $T_1$ ($T_2$) and that the two chains are isolated. The coupling between the QSCs $\hH_c$ is then introduced adiabatically and treated perturbatively within the Keldysh diagrammatic approach.~\cite{rammerBOOK07,*kamenevBOOK11} The current expectation value on the Keldysh contour reads
\beq
\label{ikeld}
I(x_1,t)=\frac{1}{2}\sum_{\rho=\pm}\sum_{\s=\pm}\langle\hI_\rho(x_1,t^\s)\rangle\equiv\frac{1}{2}\sum_{\rho=\pm}\sum_{\s=\pm}I_{\rho\s}\ ,
\eeq
where $\s=\pm$ labels the time on the forward $(+)$ and return $(-)$ contour. 
The dc noise is given by the time-averaged autocorrelation function of current at different times
\beq
\label{dcn}
S(x_1)=\sum_{\rho_1,\ \rho_2=\pm}\int dt~\langle\hI_{\rho_1}(x_1,t^-)\hI_{\rho_2}(x_1,0^+)\rangle\ .
\eeq

\subsection{Spin current}  \label{spincurrent}
We begin by computing the spin current flowing at $x=x_1$ which can be reconstructed from its components according to Eq.~\eqref{ikeld},
\beq
I_{\rho\s}=\langle \hT \hI_\rho(x_1,t^\s) e^{-\frac{i}{\hbar}\int_{c_K} dt \hH_c(t) } \rangle_0\ ,
\label{ikeld+}
\eeq
where $c_K$ indicates a time integral on the Keldysh contour and $\hT$ is the Keldysh time ordering operator. By expanding to order $\xi_\perp^2$ [i.e., $(J_c^{\perp})^2$], using Eq.~\eqref{cur} and noting that it is possible to write all quantities in terms of exponentiated boson operators via the relation $\pd_x \hvf(x,t) = \lim_{\g \to 0} (i\g)^{-1} \pd_x e^{i \g \hvf(x,t)} $, we express Eq.~\eqref{ikeld+} as
\begin{multline}
I_{\rho\s}= \frac{iu\rho}{4\p\hbar}\lim_{x \to x_1}\lim_{\g \to 0}\frac{1}{\g}\int_{c_K} dt_1\int_{c_K} dt_2\\
\times\pd_x\langle \hT e^{i \g \htvf_{2,R}(\rho x,t^\s)} \hH_c(t_1) \hH_c(t_2) \rangle_0\ .
\label{ikint}
\end{multline}
Expanding the coupling Hamiltonians using Eq.~\eqref{hc2}, we obtain
\beq
\begin{aligned}
I_{\rho\s}=& \frac{i u\rho \xi^2_\perp}{2\pi \hbar}\lim_{x \to x_1}\lim_{\g \to 0}\frac{1}{\g} \iint_{c_K}dt_1 dt_2\ e^{-i \mu(t_1 - t_2)/\hbar}\\
&\times\pd_x\langle \hT e^{i \g \htvf_{2,R}(\rho x,t^\s) - \frac{i}{K}[\htvf_{2,R}(0,t_1) - \htvf_{2,R}(0,t_2)]} \rangle_0\\
&\times\langle \hT e^{\frac{i}{K}[\htvf_{1,R}(0,t_1) - \htvf_{1,R}(0,t_2)]} \rangle_0\ , 
\end{aligned}
\label{ikint2}
\eeq
and note here that we have used the fact that the spin bias appears as a phase attached to the correlator for $\htvf_{1,R}$.

We now introduce the (equilibrium) time ordered and anti-time ordered correlation functions
\beq
\begin{aligned}
D^{++}_{\nu}(t_1,t_2)&\equiv\langle T e^{\frac{i}{K}[\htvf_{\nu,R}(0,t_1) - \htvf_{\nu,R}(0,t_2)]}\rangle_0|_{\mu=0}\\
&=\theta(t_1 - t_2) D^{-+}_\nu(t_1,t_2) + \theta(t_2 - t_1) D^{+-}_\nu(t_1,t_2)\ ,\\
D^{--}_{\nu}(t_1,t_2)&\equiv\langle \bar T e^{\frac{i}{K}[\htvf_{\nu,R}(0,t_1) - \htvf_{\nu,R}(0,t_2)]} \rangle_0|_{\mu=0}\\
&=\theta(t_2 - t_1) D^{-+}_\nu(t_1,t_2) + \theta(t_1 - t_2) D^{+-}_\nu(t_1,t_2)\ ,
\label{cor}
\end{aligned}
\eeq
where $T$ ($\bar T$) is the time (anti-time) ordering operator, $\theta$ is the Heaviside function and 
\beq
D^{\mp\pm}_{\nu}(t_1,t_2)= \curly{\frac{\frac{\pi k_B T_\nu}{u\hbar} \eta}{\sin\frac{\pi k_B T_\nu}{u \hbar}[\pm iu(t_1 - t_2) + \eta]}}^{1/K}\ .
\label{d>}
\eeq
We can then write Eq.~\eqref{ikint2} as
\beq
\begin{aligned}
I_{\rho\s}= &\frac{i u\rho \xi^2_\perp}{2 \pi \hbar}\lim_{x \to x_1}\lim_{\g \to 0}\frac{1}{\g}\sum_{\s_1,\s_2=\pm}\s_1\s_2\int_{-\infty}^{\infty} dt_1\int_{-\infty}^{\infty} dt_2\\
&\times\pd_x e^{\frac{\g}{K}\langle \hT\htvf_{2,R}(\rho x,t^\s)\htvf_{2,R}(0,t^{\s_1}_1) \rangle_0} e^{-\frac{\g}{K}\langle \hT\htvf_{2,R}(\rho x,t^\s)\htvf_{2,R}(0,t^{\s_2}_2) \rangle_0}\\
&\times D_1^{\s_1\s_2}(t_1,t_2)D^{\s_1\s_2}_2(t_1,t_2)e^{-i \mu(t_1 - t_2)/\hbar}\ ,
\label{ikrr}
\end{aligned}
\eeq
and the Keldysh contours determine the directionality of the $D_\nu$ functions. Performing the $\g\to0$ limit, we have
\beq
\begin{aligned}
I_{\rho\s}=&\lim_{x \to x_1}\frac{i u \xi_\perp^2}{2 \pi \hbar K}\sum_{\s_1,\s_2=\pm}\s_1\s_2\int_{-\infty}^{\infty} dt_1\int_{-\infty}^{\infty} dt_2\\
&\times\pd_x[f_{\rho x}^{\s\s_1}(t,t_1) - f_{\rho x}^{\s\s_2}(t,t_2)]\\
&\times D^{\s_1\s_2}_1(t_1,t_2)D^{\s_1\s_2}_2(t_1,t_2)e^{-i \mu(t_1 - t_2)/\hbar}\ ,
\label{iklong}
\end{aligned}
\eeq
where $f_{\rho x}^{\s_1\s_2}(t_1,t_2) = \langle\hT\htvf_{2,R}(\rho x,t_1^{\s_1})\htvf_{2,R}(0,t_2^{\s_2}) \rangle_0$. It is useful at this point to introduce the coordinates $\tau_0 = \frac{1}{2}(t_1 + t_2)$ and $\tau = t_1 - t_2$, and shift the resultant integrals in order to cancel out the terms in Eq.~\eqref{iklong} with $\s_1=\s_2$. We additionally see that the remaining two terms (with $\s_1=-\s_2$) can be made identical up to their phase factors by interchanging the variables $t_1\leftrightarrow t_2$ for one of the terms and noting that $D^{+-}_\nu(-\tau) = D^{-+}_\nu(\tau)$. We find that Eq.~\eqref{iklong} then reduces to
\begin{multline}
I_{\rho\s}= -\lim_{x \to x_1}\frac{u \xi_\perp^2}{\pi \hbar K}  \int_{-\infty}^{\infty} d\tau \sin{\left( \frac{\mu \tau}{\hbar} \right)} D^{-+}_1(\tau)D^{-+}_2(\tau) \\
\times \int_{-\infty}^{\infty} d\tau_0\ \pd_x\Big[ f^{\s-}_{\rho x}(0,\tau_0) - f^{\s+}_{\rho x}(0,\tau_0) \Big]\ ,
\label{ikshort}
\end{multline}
where the two integrals have been decoupled by the coordinate transformation. Lastly, we complete the $\tau_0$ integral by noting
\begin{multline}
\pd_x\Big[ f^{+-}_{\rho x}(0,\tau_0) - f^{++}_{\rho x}(0,\tau_0) \Big]=\pd_x\Big[ f^{--}_{\rho x}(0,\tau_0) - f^{-+}_{\rho x}(0,\tau_0) \Big]\\
= -\frac{2 \pi i K}{u} \theta(-\tau_0) \de(\rho x/u+\tau_0)\ ,
\label{fd}
\end{multline}
which allows us to produce the final result for this calculation
\beq
I_{\rho\s}= \frac{2 i \xi_\perp^2}{\hbar} \theta(\rho x_1) \int_{-\infty}^{\infty} d\tau \sin{\left( \frac{\mu \tau}{\hbar} \right)} D^{-+}_1(\tau)D^{-+}_2(\tau)\ .
\label{iksf}
\eeq
From Eq.~\eqref{ikeld}, we then write the final result for the bulk spin current at an arbitrary point $x_1>0$ in the drain chain as
\beq
I(x_1,t)=\frac{2 i \xi_\perp^2}{\hbar}\int_{-\infty}^{\infty} d\tau \sin{\left( \frac{\mu \tau}{\hbar} \right)} D^{-+}_1(\tau)D^{-+}_2(\tau)\ .
\label{ikf}
\eeq
The Heaviside function appearing in the expression for $I_{\rho\s}$ is a manifestation of causality: we expect no contribution from the $\rho=-$ portion in our case, as there is no coupling interaction, i.e. current tunneling into the drain chain, until the point $x=0$. 

\subsection{Dc noise}  \label{dcnoise}
The dc noise calculation proceeds similarly. We start from the general expression for the noise at $x=x_1$,
\beq
\begin{aligned}
S(x_1)&= \sum_{\rho_1,\ \rho_2=\pm}\int dt\ \langle \hT \hI_{\rho_1}(x_1,t^-) \hI_{\rho_2}(x_1,0^+) e^{-\frac{i}{\hbar} \int_{c_K}dt H_c(t)} \rangle_0\\
&\equiv\sum_{\rho_1,\ \rho_2=\pm}S_{\rho_1\rho_2}\ .
\label{nki}
\end{aligned}
\eeq
and expand this expression up to second order in $\xi_\perp$. At zeroth order, using Eqs.~\eqref{fnur} and \eqref{cur}, we obtain
\beq
\begin{aligned}
S^{(0)}_{\rho_1\rho_2}&= \int dt\ \langle \hT \hI_{\rho_1}(x_1,t^-) \hI_{\rho_2}(x_1,0^+) \rangle_0=\frac{\hbar Kk_BT_2}{2\p}\ .
\label{nki}
\end{aligned}
\eeq
So the equilibrium (Johnson-Nyquist) contribution to the dc spin current noise at $x=x_1$ is given by $S^{(0)}(x_1)=2\hbar Kk_BT_2/\p$. 

\begin{widetext}
The first non-trivial correction to this equilibrium result comes at second order in $\xi_\perp$. Representing the current operators in exponentiated form as in the spin current calculation, the nonequilibrium correction reads $S^{(2)}(x_1)=\sum_{\rho_1,\rho_2=\pm}S_{\rho_1\rho_2}^{(2)}$, where
\beq
\begin{multlined}
S^{(2)}_{\rho_1\rho_2} = \lim_{x,y \to x_1}\lim_{\g_1,\g_2 \to 0} \frac{u^2 \xi_\perp^2\rho_1\rho_2}{4 \pi^2 \g_1\g_2}\sum_{\s_1,\s_2=\pm} \s_1\s_2\\
\times\int_{-\infty}^\infty dt \int_{-\infty}^\infty dt_1 \int_{-\infty}^\infty dt_2 \ \pd_x \pd_y\langle \hT e^{i \g_1 \htvf_{2,R}(\rho_1 x,t^-)+i \g_2 \htvf_{2,R}(\rho_2 y,0^+)-\frac{i}{K} [ \htvf_{2,R}(0,t_1^{\s_1}) - \htvf_{2,R}(0,t_2^{\s_2})]}\rangle_0\langle\hT e^{\frac{i}{K} [ \htvf_{1,R}(0,t_1^{\s_1}) - \htvf_{1,R}(0,t_2^{\s_2}) ]}\rangle_0\ .
\label{nks1}
\end{multlined}
\eeq
We employ Eq.~\eqref{cor} and $f_{\rho x}^{\s_1\s_2}(t_1,t_2)$ as defined above and perform the $\g_1, \g_2 \to 0$ limits to obtain
\beq
\begin{multlined}
S^{(2)}_{\rho_1\rho_2} = \lim_{x,y \to x_1}\frac{u^2 \xi_\perp^2\rho_1\rho_2}{4 \pi^2 K^2}\int_{-\infty}^\infty dt \int_{-\infty}^\infty dt_1 \int_{-\infty}^\infty dt_2\sum_{\s_1,\s_2=\pm} \s_1\s_2\ e^{-i\mu(t_1 - t_2)/\hbar}\  D^{\s_1\s_2}_1(t_1,t_2) D^{\s_1\s_2}_2(t_1,t_2)\\
\times \pd_x\pd_y\Big[ f_{\rho_1x}^{-\s_1}(t,t_1) - f^{-\s_2}_{\rho_1x}(t,t_2) \Big]\Big[ f^{+\s_1}_{\rho_2y}(0,t_1) - f^{+\s_2}_{\rho_2y}(0,t_2) \Big]\ .
\label{nks3}
\end{multlined}
\eeq
Expanding the Keldysh contours and noting again that the contributions from $\s_1=\s_2$ vanish, we have
\beq
\begin{multlined}
S^{(2)}_{\rho_1\rho_2} = \lim_{x,y \to x_1}-\frac{u^2 \xi_\perp^2\rho_1\rho_2}{2 \pi^2 K^2}\int_{-\infty}^\infty dt \int_{-\infty}^\infty dt_1 \int_{-\infty}^\infty dt_2\cos\square{\frac{\mu(t_1 - t_2)}{\hbar}}D^{-+}_1(t_1,t_2) D^{-+}_2(t_1,t_2)\\
\times \pd_x\pd_y\Big[ f_{\rho_1x}^{--}(t,t_1) - f^{-+}_{\rho_1x}(t,t_2) \Big]\Big[ f^{+-}_{\rho_2y}(0,t_1) - f^{++}_{\rho_2y}(0,t_2) \Big]\ .
\label{nkse}
\end{multlined}
\eeq
Once again, we can transform to the coordinates $\tau_0$ and $\tau$ and find
\beq
S^{(2)}_{\rho_1\rho_2} =\lim_{x,y \to x_1}-\frac{u^2 \xi_\perp^2}{2 \pi^2 K^2}\int d\tau\cos\round{\frac{\mu\tau}{\hbar}}D_1^{-+}(\tau)D_2^{-+}(\tau)\int dt\  \Big[ f^{--}_{\rho_1x}(t,0) - f^{-+}_{\rho_1x}(t,0) \Big]\int d\tau_0\  \Big[ f^{+-}_{\rho_2y}(0,\tau_0) - f^{++}_{\rho_2y}(0,\tau_0) \Big]\ ,
\label{nks4}
\eeq
where we note that the three integrals have decoupled as in the spin current case. Equation~\eqref{fd} allows us to proceed and we obtain
\beq
S^{(2)}_{\rho_1\rho_2} = 2\xi_\perp^2 \thi(\rho_1x_1)\thi(\rho_2x_1) \int d\tau \cos{\left( \frac{\mu \tau}{\hbar} \right)} D^{-+}_1(\tau)D^{-+}_2(\tau)\ .
\label{nksf}
\eeq
\end{widetext}
Once more, as for the spin current, the Heaviside functions are manifestations of causality: we expect no nonequilibrium noise in the system until points after the tunneling site at $x = 0$. Therefore, our final results for the spin current and dc noise at $x=x_1$ are
\beq
\label{is}
I(\mu,T_1,T_2)=\frac{i(J_c^\perp)^2a^2}{2\hbar\p^2\eta^2}\int d\tau\sin\round{\frac{\mu \tau}{\hbar}}D_{1}^{-+}(\tau)D_{2}^{-+}(\tau)
\eeq
and
\beq
\begin{multlined}
\label{ss}
S(\mu,T_1,T_2)=\frac{2\hbar Kk_BT_2}{\p}\\
+\frac{(J^\perp_c)^2a^2}{2\p^2\eta^2}\int d\tau\cos\round{\frac{\mu \tau}{\hbar}}D_{1}^{-+}(\tau)D_{2}^{-+}(\tau)\ ,
\end{multlined}
\eeq
respectively.

While we have presented a Keldysh calculation for the spin current and noise at $x=x_1$, Eqs.~\eqref{is} and \eqref{ss} could have been obtained instead by computing the tunneling spin current and its noise {\em at the coupling site $x=0$}. We have verified that this latter calculation results in a current that is identical to Eq.~\eqref{is} and a noise that is identical to the second term in Eq.~\eqref{ss}. This outcome is physically sensible. As apparent from Eq.~\eqref{hnus2}, the QSCs are modeled as an essentially free boson gas. Therefore, nonequilibrium disturbances produced at the upstream end should remain unmodified as they propagate downstream to $x=x_1$. In particular, tunneling spin current at the coupling site $x=0$ should be identical to the current downstream. Moreover, any additional noise generated at the left end of the drain QSC should propagate downstream undisturbed. We will be using this fact in the proceeding magnon transport comparison calculation.

\begin{figure}[t]
\includegraphics[width=0.9\linewidth]{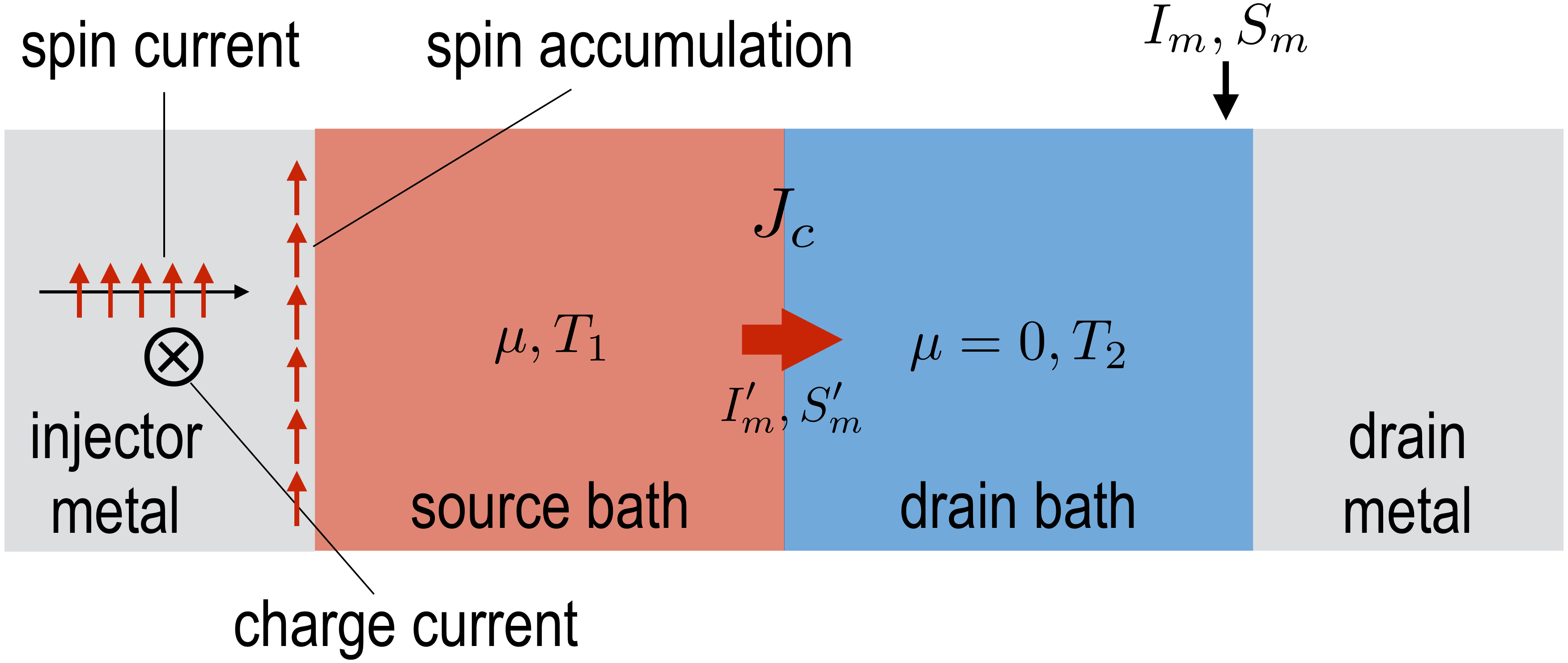}
\caption{(Color online) Depiction of proposed magnonic system. Two coupled magnon baths are held at temperatures $T_1$ and $T_2$ and chemical potentials $\mu_1$ and $\mu_2$, respectively.} 
\label{fig3}
\end{figure}


\subsection{Spin transport between two magnon baths}
\label{mag}
Equations~\eqref{is} and \eqref{ss} can now be contrasted to the case of spin transport across two coupled magnon baths (see Fig.~\ref{fig3}). In similar spirit to the QSC calculation, our interest here is the spin current $I_m$ flowing at the spatial point just left of the drain metal (indicated by the black arrow) and its dc noise $S_m$, as these are the physical quantities that may be electrically detected by the metal. We consider both raising the magnon chemical potential in the source bath above zero and/or the spin Seebeck effect,~\cite{maekawaBOOK12} in which spin current is generated by a temperature difference between the two baths. SHEs can be utilized to transfer spin angular momentum from the injector metal into the source magnon bath and raise the magnon chemical potential in the latter;~\cite{cornelissenNATP15,goennenweinAPL15,cornelissenPRB16,shanPRB16} the injection process is analogous to the spin injection process discussed in the context of the QSC setup (see Sec.~\ref{model}). 

In the absence of the bath coupling, we assume that magnon bath $\nu$ is thermalized to a distribution function $n_\nu(\W)=\{\exp[\be_\nu(\hbar\W-\mu_\nu)]-1\}^{-1}$, where $\nu=1,2$ labels the two magnon baths, $\be_\nu=(k_BT_\nu)^{-1}$ is the inverse temperature, and $\mu_\nu$ is the magnon chemical potential; we set $\mu_1=\mu$ and $\mu_2=0$ throughout unless otherwise stated.

We work with two identical spin-$s$ Heisenberg ferromagnetic insulators (with $N_x\times N_y\times N_z$ cubic lattice structures) in a uniform magnetic field along the $z$ axis 
\beq
\hH^m_{\nu}=-\frac{J}{2}\sum_{\bj,\bde}\hbS_{\nu,\bj}\cdot\hbS_{\nu,\bj+\bde}+H\sum_{\bj}\hS^z_{\nu,\bj}\ ,
\label{hm}
\eeq
where $J$ is the exchange coupling, $H$ is the Zeeman energy due to an external magnetic field along the $z$ axis, $\bj$ labels sites of the lattice, and $\bde$ labels all nearest neighbor sites (this model has been applied to, e.g., YIG with $s\approx14$ and lattice constant $a\approx 12\AA$~\cite{rueckriegelPRB14}). Assuming $s\gg1$, Eq.~\eqref{hm} maps to an essentially non-interacting boson model via $\hS^-_{\nu,\bj}\approx\sqrt{2s}\hb_{\nu,\bj}$ and $\hS^z_{\nu,\bj}=\hb_{\nu,\bj}^\dag\hb_{\nu,\bj}-s$.~\cite{holsteinPR40} The system is comprised of two semi-infinite systems coupled along the $x=0$ plane, thus we consider the boundary and so diagonalize Eq.~\eqref{hm} via the Fourier transform
\beq
\label{ftm}
\hb_{\nu,\bj}=\sqrt{\frac{2}{N_xN_yN_z}}\sum_{\bk}e^{ik_yj_ya+ik_zj_za}\cos(k_xj_xa) \hb_{\nu,\bk}\ ,
\eeq
which is appropriate for a zero-flux boundary condition at the interface; the momenta are $k_x=\p n_x/N_xa$ and $k_{y,z}=2\p n_{y,z}/N_{y,z}a$, with $n_i=1,\dots,N_i$. We then arrive at $\hH^m_\nu=\sum_{\bk}\ve_\bk\hb^\dag_{\nu,\bk}\hb_{\nu,\bk}$ with magnon dispersion $\ve_\bk=-2Js[\cos(k_xa)+\cos(k_ya)+\cos(k_za)-3]+H$. 
Finally, we use a general interfacial exchange coupling 
\beq
\hH^{m}_{c}=s\sum_{\bk,\bp}J_{c\bk\bp}b^{\dag}_{\bk,1}b_{\bp,2}+h.c.\ . 
\eeq

We note that for $s\gg1$ as assumed here, Eq.~\eqref{hm} maps to an essentially free boson model. Therefore, as discussed at the end of Sec.~\ref{dcnoise}, tunneling spin current across the two magnon baths and its noise (denoted by $I'_m$ and $S'_m$, respectively, in Fig.~\ref{fig3}) enter the drain bath at $x=0$ and should propagate undisturbed down to the drain metal, where they can be detected. We therefore expect here that $I_m=I'_m$ and $S_m=S^{(0)}_m+S'_m$, where $S^{(0)}_m$ is the equilibrium (Johnson-Nyquist) spin current noise present in the drain magnon bath even in the absence of $J_c$. Here, we present calculations for $I'_m$ and $S'_m$.

The operator for spin current tunneling across the two baths can be defined as the total spin leaving bath 1, 
\beq
\hI'_m(t)=-\hbar\pd_t\sum_{\bj}\hS^{z}_{1,\bj}\ .
\eeq 
Computing the nonequilibrium expectation value of $\hI'_m$ to lowest non-trivial order in $J_c$, we may collapse the Keldysh contour and represent the result more compactly as
\beq
\label{icol}
I'_m=-\frac{i}{\hbar}\int dt~\thi(-t) \langle[\hI'_m(0),\hH^m_c(t)]\rangle_0\ ,
\eeq
where $\langle\rangle_0$ depicts an equilibrium average with respect to the above-mentioned thermal states. Again, to lowest non-trivial order in $J_c$, the dc noise is given by the time-averaged autocorrelation function of current at different times, i.e.,
\beq
\label{dcn}
S'_m=\int dt~\langle\hI'_m(t)\hI'_m(0)\rangle_0\ .
\eeq
Denoting the ``greater" and ``lesser" Green functions for the magnons via $G^>_{\bk,\nu}(t) = -i\langle b_{\bk,\nu}(t)b_{\bk,\nu}^\dag (0) \rangle_0$ and $G_{\bk,\nu}^<(t) = -i\langle b_{\bk,\nu}^\dag (0) b_{\bk,\nu}(t) \rangle_0$, we then obtain
\begin{align}
\label{imo}
I'_m&=-\frac{J_c^2s^2}{2\pi\hbar}\int d\Omega\sum_{\bk,\bp}\Big\{ G^<_{\bk , 1}(\Omega) G^>_{\bp , 2}(\Omega) - G^>_{\bk , 1}(\Omega) G^<_{\bp , 2}(\Omega) \Big\}\ , \\
\label{smo}
S'_m&=-\frac{J_c^2 s^2}{2\pi} \int d\Omega \sum_{\bk , \bp} \Big\{ G^<_{\bk , 1}(\Omega) G^>_{\bp , 2}(\Omega) + G^>_{\bk , 1}(\Omega) G^<_{\bp , 2}(\Omega) \Big\}\ ,
\end{align}
where we have taken $J_c\equiv J_{c\bk\bp}$. The Fourier transformed Green functions are $G^<_{\bk,\nu}(\Omega) = -2 \pi in_\nu(\Omega) \de(\Omega - \varepsilon_{\bk}/\hbar)$ and $G^>_{\bk,\nu} (\W)= -2 \pi i[1+n_\nu(\Omega)]\de(\Omega - \varepsilon_\bk/\hbar)$, where $n_\nu$ is the Bose distribution defined above and $\ve_\bk \approx Js(ka)^2 + H$ is the magnon dispersion. 

Then the spin current and its noise downstream near the drain metal should read
\begin{align}
\label{im}
&I_m(\mu,T_1,T_2)=\frac{1}{\hbar}\int_{H/\hbar}d\W\ g(\W)[n_1(\W) - n_2 (\W)]\ ,\\
&S_m(\mu,T_1,T_2)=S^{(0)}_m\\
\label{sm}
&\quad+\int_{H/\hbar}d\W\ g(\W)[n_1(\W)+n_2(\W)+2n_1(\W)n_2(\W)]\ ,
\end{align}
where $g(\W)= (J_c\hbar N_xN_yN_z)^2(\hbar\W-H)/8\p^3J^3s$ encodes the magnon tunneling density of states. The exact expression for the equilibrium noise component $S^{(0)}_m$ will not be essential in the remainder of the discussion.

\section{Discussion}
\label{disc}
\subsection{Spin Fano factor}
\label{fano}
\begin{figure}[t]
\includegraphics[width=0.8\linewidth]{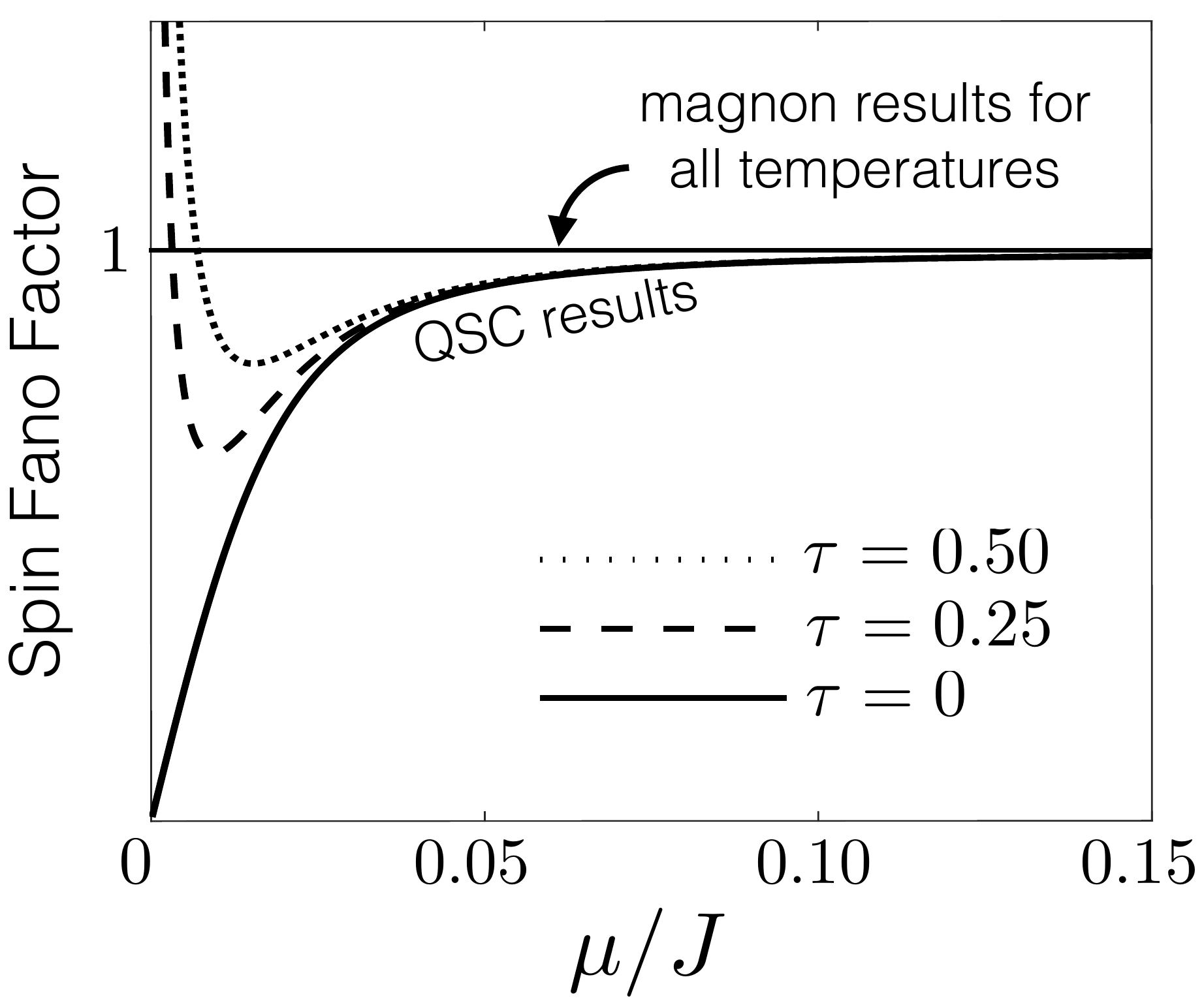}
\caption{Spin Fano factor in the QSC and magnon cases plotted as a function of the chemical bias $\mu\equiv\mu_1$. Here, $\tau=(T_1-T_0)/T_0$, and we set $\mu_2=0$ and $T_2\equiv T_0$ in both systems.} 
\label{fig5}
\end{figure}

We begin by fixing the temperature of subsystem $\nu=2$ to $k_BT_2/J=0.004$ in the QSC case (e.g., $T_2\sim10$~K for Sr$_2$CuO$_3$ with $J\approx2000$~K~\cite{hirobeNATP17}) and $k_BT_2/Js=0.08$ in the magnon case (e.g., $T_2\sim4$~K for YIG~\cite{rueckriegelPRB14}). We define the dimensionless thermal bias $\tau\equiv (T_1-T_2)/T_2$ in both cases and the {\em nonequilibrium noise} $S_{\rm neq}(\mu,T_1,T_2)\equiv S(\mu,T_1,T_2)-S_{\rm eq}$ [$S_{m,\rm neq}(\mu,T_1,T_2)\equiv S_m(\mu,T_1,T_2)-S_{m,\rm eq}$ for the magnon case], where $S_{\rm eq}\equiv S(\mu=\tau=0)$ [$S_{m,\rm eq}\equiv S_m(\mu=\tau=0)$] is the background (thermal) noise in the absence of any bias. Fig.~\ref{fig5} then depicts the {\em spin Fano factor}, defined as $F\equiv S_{\rm neq}/\hbar I$ and $F_m\equiv S_{m,\rm neq}/\hbar I_m$ as a function of the chemical bias $\mu$ for various temperature biases $\tau$. 


From Eq.~\eqref{im}, we see that a finite magnon current $I_m$ can be generated with either a finite $\mu$ or a finite $\tau$, and we find $F_m=1$ for any $\mu$ and/or $\tau$. The spin Fano factor defined here corresponds only to the nonequilibrium contribution to the spin current noise. Therefore, $F_m=1$ reflects the (uncorrelated) Poissonian tunneling of magnons, each carrying a spin quantum of $\hbar$, generated by the nonequilibrium biases.~\footnote{Ref.~\onlinecite{kamraPRL16} has calculated shot noise in the spin current generated via spin pumping with a mono-domain ferromagnet and reported super-Poissonian shot noise resulting from dipolar interactions.} 

The QSC spin Fano factor behaves markedly different. As $\mu$ increases, i.e., enters the regime $\mu\gg k_BT_2$, the QSC spin Fano factor approaches 1, the same value as the magnon spin Fano factor. This shows that for large biases (i.e., in the shot limit), spin current across the two QSCs is mediated by a Poissonian tunneling of spin-1 excitations, consistent with the inter-chain exchange coupling Eq.~\eqref{hc} which transfers spin-1 excitations across the QSCs. However, the QSC spin Fano factor vanishes to $0$ as $\mu\rightarrow0$ for $\tau = 0$ and diverges for any $\tau>0$ as $\mu\rightarrow0$. The vanishing of $F$ can be understood by noticing from Eqs.~\eqref{is} and \eqref{ss} that $I\propto\mu$ and $S_{\rm neq}\propto\mu^2$ for $\tau=0$ as $\mu\rightarrow0$. Physical speaking, this points to the fact that at $\ta=0$ excess spin noise cannot depend on which chain is biased, unlike spin current which must. Ultimately the spin Fano factor vanishes like $F\propto\mu$ in the absence of temperature bias.

The divergence of $F$ for $\tau>0$ can be seen formally in Eq.~\eqref{is}, where the effects of thermal and chemical biases completely factorize and $\mu$ appears only inside the sine prefactor. The spin current $I$ decreases to zero as $\mu\rightarrow0$ for any $\tau$ while the excess noise $S_{\rm neq}$ remains finite for all $\mu$ and $\tau>0$ [see Eq.~\eqref{ss}], so $F$ diverges as $\mu\rightarrow0$ for all $\tau>0$. Within the gaussian Luttinger model description for the QSCs used here, a thermal bias alone does not generate a net spin current across the chains, and a nonzero chemical bias $\mu>0$ is required for a net spin flow between the chains. Interpreted physically, what we find is the exposure of the underlying fermionic statistics of quasiparticle exchange across the weak link. The exchange coupling, Eq.~\eqref{hc}, represents hopping of Jordan-Wigner fermions. When $\mu=0$, as the fermionic spectrum is linearized and the density of states is thus constant at the fermi points, thermal bias on one edge leads to no net current across the weak link whereas excess noise still arises. This physical interpretation is represented in Fig.~\ref{fig5} as divergence of the Fano factor for $\tau > 0$ as $\mu \to 0$.




In Sec.~\ref{fqh}, we noted an exact analogy between the current problem and the problem of stochastic tunneling of electrons between two single fractional quantum Hall edge channels. In the latter problem, the factorization of the thermal and chemical (i.e., voltage) biases in the tunneling charge current has been obtained and is well-known.~\cite{kaneBOOK07,chamonPRL93,kanePRL94}

\subsection{Manifestation of Pauli blockade in noise}
\label{pbn}
The $xxz$ quantum antiferromagnetic chains described by Eq.~\eqref{hnu} can be mapped to a lattice model of interacting fermions via the Jordan-Wigner transformation~\cite{jordanZP28}
\beq
\label{jwt}
\hS_j^z=\hc_j^{\dagger}\hc_j-\frac{1}{2}\ ,\quad\hS_j^-=\hc_j\cos\round{\p\sum_{l<j}\hc_l^{\dagger}\hc_l}=\round{\hS_j^+}^{\dagger}\ ,
\eeq
with $\hS^\pm_j=\hS^x_j\pm i\hS^y_j$. This mapping to the fermion model reveals a certain resemblance between the spin-1/2 operators and fermions where the absence (presence) of a fermion on site $j$ corresponds to a state with $S^z_j=-1/2$ ($S^z_j=1/2$). In particular, once a spin-1 excitation is injected into the drain chain at $j=0$ from the source chain, a second spin-1 excitation cannot be injected into the same site; this leads to a partial blockade of spin transport across the chains, analogous to transport blockade due to Pauli's exclusion principle observed in electron transport.~\cite{onoSCI02}

\begin{figure}[t]
\includegraphics[width=0.75\linewidth]{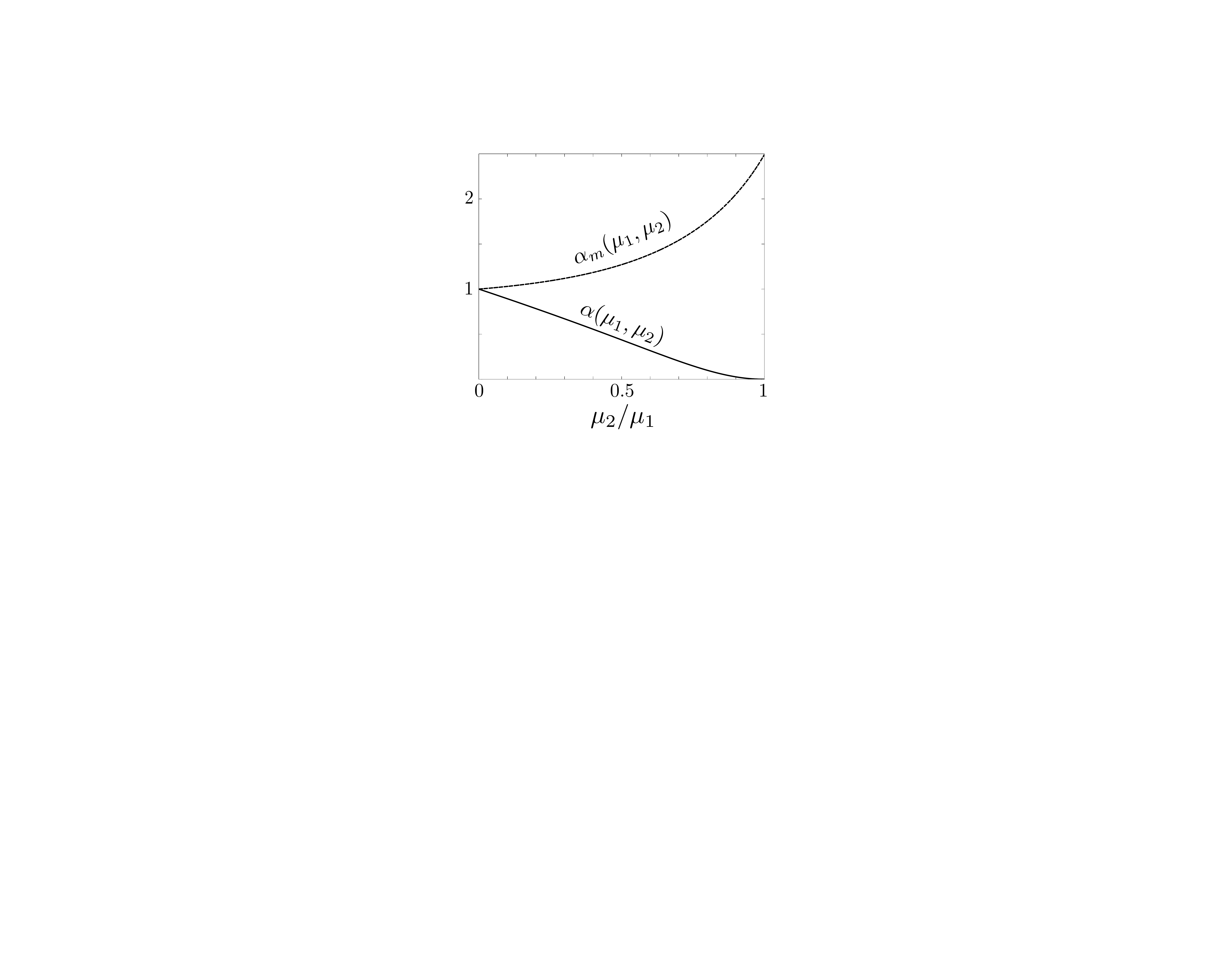}
\caption{Excess dc spin current noise as a function of $\mu_2/\mu_1$ for the QSC case (solid line) and the magnon case (dashed line) with $T_1=T_2=T_0$. See the main text for the definitions of $\al(\mu_1,\mu_2)$ and $\al_m(\mu_1,\mu_2)$. The suppression obtained in the QSC case is a manifestation of Pauli blockade physics.}
\label{fig6}
\end{figure}

A direct consequence of this Pauli blockade physics in QSCs can be obtained in the dc noise, the physical origin of which exactly resembles the Pauli blockade picture developed for charge fluctuation suppression between two edge states of a fractional quantum Hall liquid coupled at a quantum point contact.~\cite{takeiPRB15} To elaborate on this point, let us now consider a case when both QSCs are chemically biased such that spin currents impinge the junction from both ends in Fig.~\ref{fig1}. If we denote the spin chemical potentials for the two QSCs as $\mu_1$ and $\mu_2$, the nonequilibrium dc noise is modified to
\beq
\begin{multlined}
S_{\rm neq}(\mu_1,\mu_2,T_1,T_2)\\
=\frac{(J_c^\perp)^2a^2}{2\p^2\eta^2}\int dt\curly{\cos\square{\frac{(\mu_1-\mu_2)t}{\hbar}}-1}D^{-+}_1(t)D^{-+}_2(t)\ .
\end{multlined}
\eeq
We now fix the temperatures of both QSCs to the same value, i.e., $T_1=T_2\equiv T_0$, and define 
\beq
\alpha(\mu_1,\mu_2) \equiv \frac{S_{\rm neq}(\mu_1,\mu_2,T_0,T_0)}{S_{\rm neq}(\mu_1,0,T_0,T_0)}
\eeq 
as the ratio of the nonequilibrium noises when both QSCs are chemically biased to that when only one of the QSCs is biased. Fig.~\ref{fig6} is then produced by sweeping $\mu_2$ from $0$ to $\mu_1$ while keeping $\mu_1$ fixed. In the QSC system (denoted by the solid line), the nonequilibrium noise exhibits suppression as $\mu_2\rightarrow\mu_1$ and it vanishes at $\mu_2=\mu_1$. This noise reduction can be attributed to Pauli blockade, which suppresses the phase space for the scattering of (fermionic) spin-1 excitations at the QSC junction. 

A starkly contrasting behavior is predicted for the magnon setup. Here, we consider a possibility of both magnon baths having finite chemical potentials $\mu_1$ and $\mu_2$, and we define the ratio 
\beq
\alpha_m(\mu_1,\mu_2) \equiv \frac{S_{m,\rm neq}(\mu_1,\mu_2,T_0,T_0)}{S_{m,\rm neq}(\mu_1,0,T_0,T_0)}\ ,
\eeq 
which again is the ratio of the nonequilibrium noises when both magnon baths are chemically biased to that when only one bath is biased. In the magnon case, introducing $\mu_2$ results in more noise, i.e., $\al_m>1$ for $\mu_2>0$ exhibiting no signature of Pauli blockade, which is a feature unique to fermionic excitations.


\section{Experimental extraction of spin Fano factor}
\label{exp}

\subsection{QSC scenario}

\begin{figure}[t]
\includegraphics[width=\linewidth]{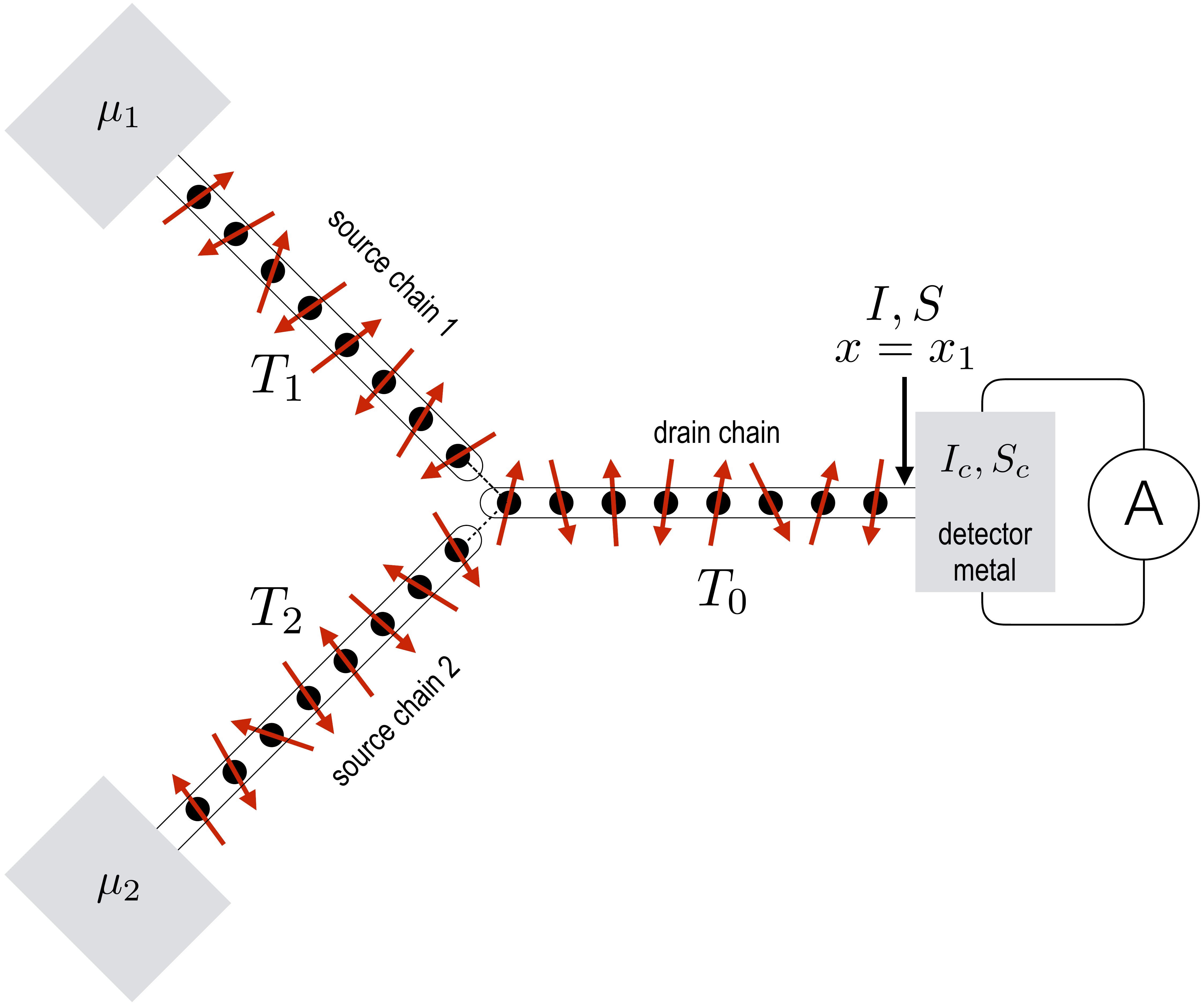}
\caption{(Color online) Depiction of the proposed configuration for extracting the spin Fano factor in the QSC case. Two source chains are individually coupled to a third drain chain at one point. The chains are held at temperatures $T_1$, $T_2$, and $T_0$ respectively, and chemical biases $\mu_1$, $\mu_2$ in the source chains only.}
\label{fig7}
\end{figure}

The QSC spin Fano factor may be experimentally detected using a Y-junction setup shown in Fig.~\ref{fig7}. In this setup, the drain QSC is weakly exchange coupled to two source QSCs at its left end and to a metal with strong spin-orbit coupling at its right; the source chains 1 and 2 are chemically biased by spin chemical potentials $\mu_1$ and $\mu_2$, respectively. We assume that the two source QSCs are identical and that they are exchange coupled with equal strength to the drain chain; departures from this symmetric condition should not affect the following discussion at the qualitative level. The temperatures of source chain 1, source chain 2 and the drain chain are denoted by $T_1$, $T_2$ and $T_0$, respectively.

To lowest order in the source-drain coupling, the spin current $I$ at $x=x_1$ reads
\beq
\begin{multlined}
I=I(\mu_1,T_1,T_0)+I(\mu_2,T_2,T_0)\ ,
\end{multlined}
\eeq
and its dc noise is given by
\beq
\begin{aligned}
S(\mu_1,\mu_2,T_1,T_2)&=\frac{2\hbar Kk_BT_0}{\p}\\
&+\frac{(J_c^\perp)^2a^2}{2\p^2\eta^2}\int dt\cos\round{\frac{\mu_1t}{\hbar}}D_{1}^{-+}(t)D_{0}^{-+}(t)\\
&+\frac{(J_c^\perp)^2a^2}{2\p^2\eta^2}\int dt\cos\round{\frac{\mu_2t}{\hbar}}D_{2}^{-+}(t)D_{0}^{-+}(t)\\
&\equiv S^{(0)}+S^{(2)}(\mu_1,T_1,T_0)+S^{(2)}(\mu_2,T_2,T_0)\ .
\end{aligned}
\eeq
Spin current fluctuations in the drain QSC at $x=x_1$ should generate pure spin current fluctuations inside the adjacent metal due to the coupling between the two systems. The latter pure spin current noise should convert into noise in the transverse charge current via the inverse SHE and should be detectable using an ammeter as illustrated in Fig.~\ref{fig7}. 

An electrical noise measurement is first made when the two source chains are unbiased, i.e., $\mu_1=\mu_2=0$, and $T_1=T_2=T_0$. We denote the electrical noise measured in this configuration with $S_{c,\rm eq}$.

The charge noise is then measured in a second configuration in which the source chains are asymmetrically biased, i.e., $\mu_1=-\mu_2$, such that $I$ remains zero. We denote the noise measured in this configuration with $S_c$ and the excess nonequilibrium noise as $S_{c,\rm neq}\equiv S_c-S_{c,\rm eq}$. The excess noise in the second configuration should arise solely from the nonequilibrium spin current noise generated at the coupling site. Using the notation of Sec.~\ref{fano}, this nonequilibrium spin current noise is given by $S_{\rm neq}(\mu_1,T,T_0)+S_{\rm neq}(-\mu_1,T,T_0)=2S_{\rm neq}(\mu_1,T,T_0)$, where we have assumed $T\equiv T_1=T_2$ due to the above symmetry condition. It is therefore reasonable to assume that the excess electrical noise is proportional to the excess spin current noise, i.e., $S_{c,\rm neq}=\Theta_SS_{\rm neq}(\mu_1,T,T_0)$, where $\Theta_S$ is some spin-to-charge noise conversion constant. The temperature of the source chains $T$ may be elevated above that of the drain chain $T_0$ because the generation of finite $\mu_1$ and $\mu_2$ requires charge currents in the injector metals and may cause Joule heating.



Lastly, we turn $\mu_2$ off while keeping $\mu_1$ at the same value as in the second configuration. There is now a net spin current $I(\mu_1,T,T_0)$ flowing into the drain QSC that will convert into a net charge current via the inverse SHE inside the detector metal. It is reasonable to assume here that this generated charge current is proportional to the generated spin current $|I_c|=\Theta_I|I(\mu_1,T,T_0)|$, where $\Theta_I$ is some spin-to-charge current conversion constant.

Finally, the electrical Fano factor can be computed from the experimental readings, and we find that
\beq
F_c=\frac{S_{c,\rm neq}}{|I_c|}=\frac{\Theta_S}{\Theta_I}\frac{S_{\rm neq}(\mu_1,T,T_0)}{|I(\mu_1,T,T_0)|}=\frac{\hbar\Theta_S}{\Theta_I}F(\mu_1,T,T_0)\ .
\eeq 
From Fig.~\ref{fig5}, we know that for any $T$ and $T_0$, the spin Fano factor $F$ approaches 1 as $\mu_1$ increases. In the large chemical bias regime where $F$ approaches 1, the ratio of the unknown prefactors can be experimentally extracted via
\beq
\frac{\Theta_S}{\Theta_I}=\frac{F_c}{\hbar}\ ,
\eeq
thus allowing one to obtain the spin Fano factor $F$ for all $\mu_1$ values.

If the temperatures of the source QSCs and the drain QSC are uniform, i.e., $T=T_0$, $F_c$ should display a vanishing behavior as $\mu_1\to0$. However, if Joule heating results in $T>T_0$, a thermal bias exists in addition to the chemical bias and we expect $F$ to exhibit a diverging behavior shown in Fig.~\ref{fig5}. Therefore, the spin Fano factor can be used to distinguish between the presence and absence of Joule heating in the source QSC due to charge currents in the injector metals.

\begin{figure}[t]
\includegraphics[width=\linewidth]{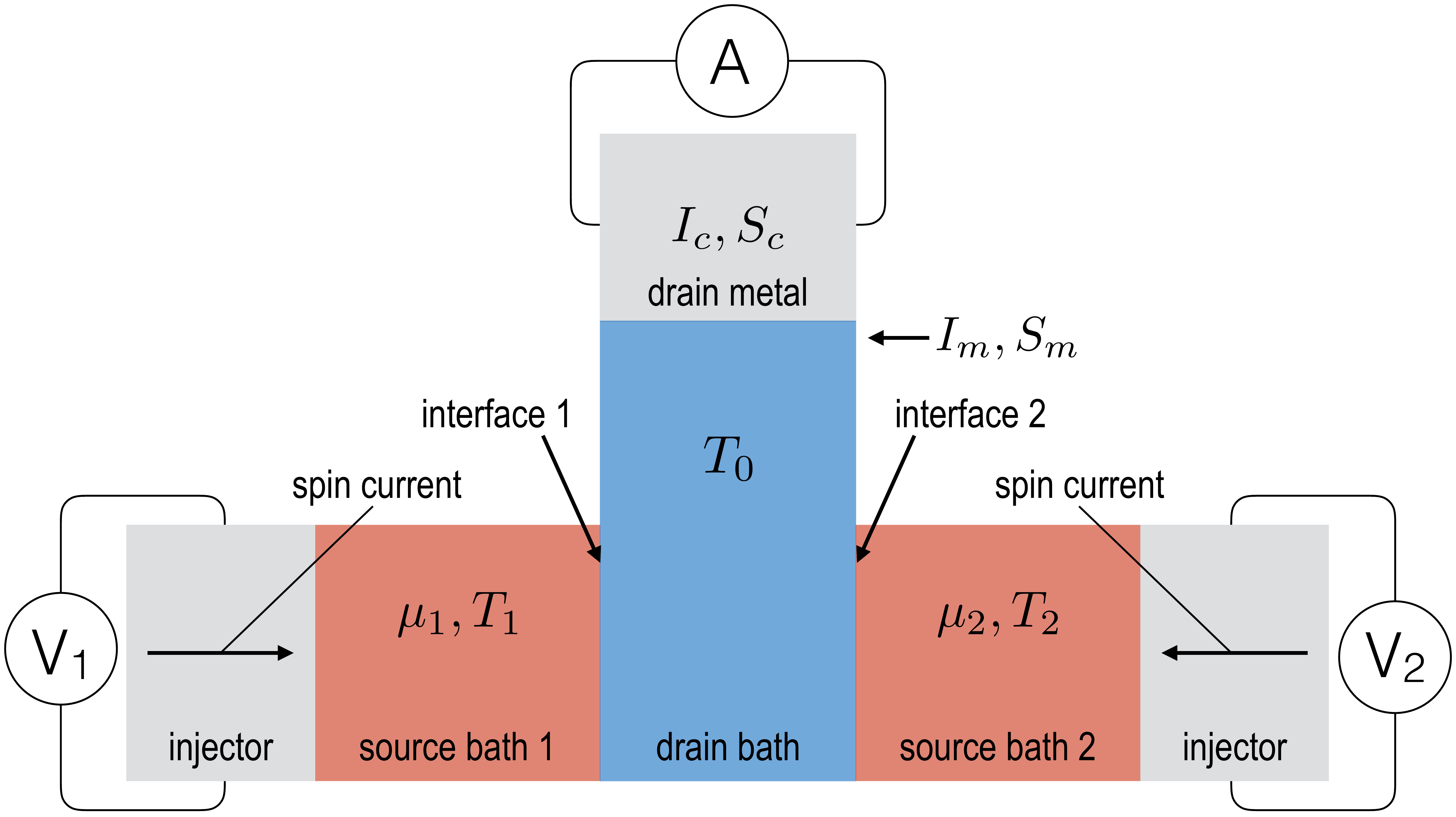}
\caption{(Color online) Depiction of the proposed configuration for extracting the spin Fano factor in the magnon case. Two source baths are individually coupled to a third drain bath. The baths are held at temperatures $T_1$, $T_2$, and $T_0$ respectively, and chemical biases $\mu_1$, $\mu_2$ in the source baths alone.}
\label{fig8}
\end{figure}

\subsection{Magnon scenario}

The spin Fano factor for the magnon scenario can be extracted in a setup similar to Fig.~\ref{fig7}. We consider the drain magnon bath coupled to two source magnon baths in a T-shaped setup as shown in Fig.~\ref{fig8}. The injection of spin angular momentum into the source magnon baths are facilitated by charge currents and SHE in the respective injector metals. An important consequence of this injection process is to generate finite chemical potentials $\mu_1$ and $\mu_2$ for the source magnon baths. We assume the symmetry condition (as in the QSC setup), in which the two source magnon baths are identical and they couple to the drain magnon bath with an equal strength. The temperatures of magnon bath 1, magnon bath 2 and the drain magnon bath are denoted by $T_1$, $T_2$ and $T_0$, respectively.

Electrical noise is first measured when the two source magnon baths are unbiased, i.e., $\mu_1=\mu_2=0$, and $T_1=T_2=T_0$. We denote the electrical noise measured in this configuration with $S_{c,\rm eq}$.

In the second configuration, the source baths are oppositely biased, i.e., $\mu_1>0$ and $\mu_2<0$, such that no net magnon spin current is detected at the detector metal, i.e., $I_m(\mu_1,T_1,T_0)=-I_m(\mu_2,T_2,T_0)$ such that $I_m=0$. We again denote the electrical noise measured in this second configuration with $S_c$ and the excess noise as $S_{c,\rm neq}\equiv S_c-S_{c,\rm eq}$. As in the QSC scenario, the excess noise $S_{c,\rm neq}$ should arise solely from the excess spin current noises that are generated at interfaces 1 and 2 and have propagated to the detector metal. This excess noise in the spin current is given by $S_{m,\rm neq}(\mu_1,T_1,T_0)+S_{m,\rm neq}(\mu_2,T_2,T_0)$. It is then reasonable to assume that the excess electrical noise measured by the detector metal is proportional to this excess spin current noise, i.e., $S_{c,\rm neq}\propto S_{m,\rm neq}(\mu_1,T_1,T_0)+S_{m,\rm neq}(\mu_2,T_2,T_0)$. The temperature of the source baths may once again be elevated above that of the drain bath $T_0$ because charge currents in the injector metals may cause Joule heating in the magnon baths. 


Once $\mu_2$ is turned off while keeping $\mu_1$ at the same value as above, a net spin current $I_m(\mu_1,T_1,T_0)$ flows into the drain bath. It is reasonable to assume that the charge current generated at the detector metal obeys $I_c\propto I_m(\mu_1,T_1,T_0)$.

Taking the ratio $F_c=S_{c,\rm neq}/|I_c|$, we find that 
\beq
\begin{aligned}
F_c&\propto\frac{S_{m,\rm neq}(\mu_1,T_1,T_0)+S_{m,\rm neq}(\mu_2,T_2,T_0)}{|I_m(\mu_1,T_1,T_0)|}\\
&=\frac{S_{m,\rm neq}(\mu_1,T_1,T_0)}{|I_m(\mu_1,T_1,T_0)|}+\frac{S_{m,\rm neq}(\mu_2,T_2,T_0)}{|I_m(\mu_2,T_2,T_0)|}\propto F_m\ ,
\end{aligned}
\eeq
where $F_m$ is shown in Fig.~\ref{fig5}. Here, unlike the QSC case, regardless of the temperatures and relative biases of the magnon baths, $F_c$ should remain constant. Therefore, the contrasting behavior between the QSC and magnon spin Fano factors, as plotted in Fig.~\ref{fig5}, should be electrically measurable via inverse SHE signals.

\begin{figure}[t]
\includegraphics[width=0.7\linewidth]{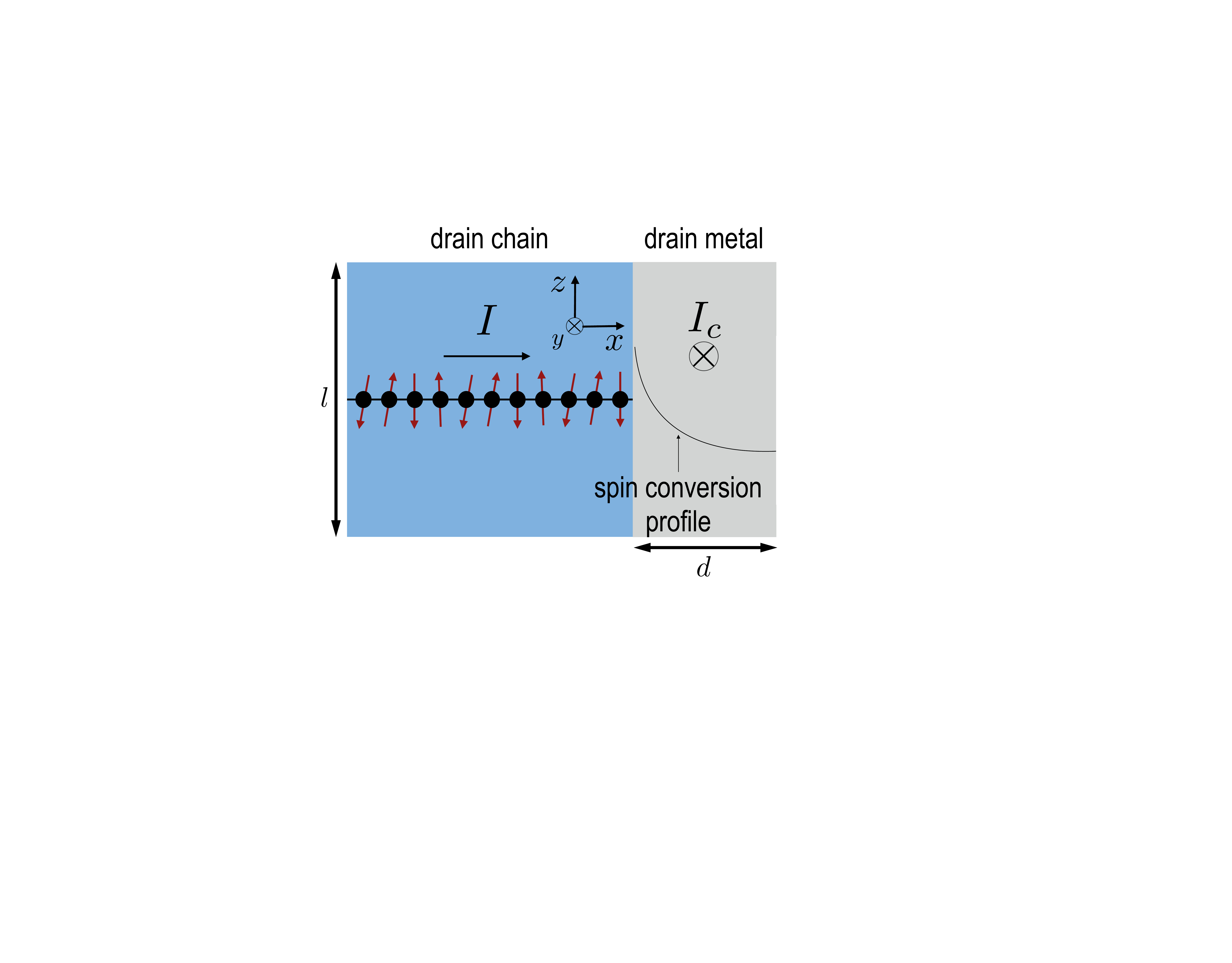}
\caption{(Color online) Depiction of spin to charge current conversion between a QSC and normal metal contact. We extract an approximation for the size of the conversion effect between the two materials from this geometry. Here, $l$ is the thickness of the interface and $d$ is the width of the metal contact.}
\label{fig9}
\end{figure}

\subsection{Estimate of noise signal strength}

We derive an estimate for the magnitude of the noise effect measurable via conversion in a strongly spin-orbit coupled metal spin drain (e.g., Pt). Fig.~\ref{fig9} shows the geometry of the contact, which is standard in recent two terminal spin chain experiments, for instance in Ref.~\onlinecite{hirobeNATP17} (using spin chain material Sr$_2$CuO$_3$). If we consider $N$ QSCs attached laterally per unit area, we may write the total spin current density impinging on the normal metal as $NI(t)$, where $I(t)$ is the spin current due to a single QSC. For a normal metal of thickness $d$ and spin diffusion length $\la$, we may write the spin current density profile across the normal metal as $j_s(x,t) = NI(t) \sinh{\left( \frac{d-x}{\la} \right)}/\sinh{\left( \frac{d}{\la} \right)}$,~\cite{mosendzPRB10} assuming that the spin current $I(t)$ [see Eq.~\eqref{is}] arriving at the interface fully penetrates into the metal and a boundary condition of vanishing spin current at the outer edge of the sink material. Modeling spin to charge conversion via the SHE, characterized by the spin Hall angle $\Theta$, we derive the charge current density in the normal metal as $j_c(x,t) = \Theta \frac{2e}{\hbar} j_s(x,t)$. We then find the total charge current flowing in the metal by integrating over the cross-section 
\beq
I_c(t)= \Theta \frac{2e}{\hbar} N l \la I(t) \tanh{\left( \frac{d}{2 \la} \right)}\ ,
\eeq
where $l$ is the height of the interface area (see Fig.~\ref{fig9}), and characterize the associated dc charge noise $S_c=\int dt\ \langle I_c(t)I_c(0) \rangle$ as
\beq
S_c=\left[ \Theta \frac{2e}{\hbar} N l \la \tanh{\left( \frac{d}{2 \la} \right)} \right]^2 S. \label{curn}
\eeq

We determine $N$ from the documented values for the lattice spacings of Sr$_2$CuO$_3$,~\cite{johnstonBOOK97} use $\Theta=0.1$ and take $\la = 2~\mathrm{nm}$, $d = 7~\mathrm{nm}$, and $l = 1~\mathrm{mm}$. Using the known properties of Pt, we estimate the magnitude of the fluctuations as $S_c \sim10^{-16}~V^2s$ in the given configuration. This calculation assumes a temperature of $T=20$~K, again following Ref.~\onlinecite{hirobeNATP17}, which results in background Johnson-Nyquist noise of $S_{JN} \sim 10^{-14}~V^2s$ in the contact. Voltage noise measurements of order $10^{-20}~V^2s$ have been reported in, e.g., Ref.~\onlinecite{kamraPRB14}. We thus believe the Fano factor results of Sec.~\ref{fano} should be measurable with existing equipment and techniques.

\section{Conclusion} \label{conc}

We have considered two spin systems: two semi-infinite QSCs that generate noise across a weak coupling, and two semi-infinite magnon baths that also generate noise across a weak coupling. In either case, we have derived the bulk spin current and bulk spin current noise at a point close to a measurement metal contact with strong spin-orbit coupling. Our analysis shows that it should be possible to differentiate the systems using a quantity known as the \textit{spin Fano factor}, where Pauli blocking and concomitant current suppression in the QSC case result in dramatically different behavior than for the same quantity in the magnon scenario. Additionally, we show that Pauli blocking has signatures directly accessible in the noise. These results exhibit the fermionic nature of the spin-$1/2$ operator in the QSC.

We then propose an experimental method by which to extricate the spin Fano factor and therefore experimentally compare the two systems. In so doing we assumed proportional relationships between the excess spin current noise and the excess charge noise, and the spin current and charge current in the metallic contacts, i.e., both $S_{c,\rm neq}=\Theta_SS_{\rm neq}(\mu_1,T,T_0)$ and $|I_c|=\Theta_I|I(\mu_1,T,T_0)|$. However, it would be desirable to develop a microscopic model for the conversion between these quantities thereby elucidating the regime of validity of the assumption and providing microscopic determinations of $\Theta_S$ and $\Theta_I$. Additionally, we have not extended this work to careful examination of higher order effects, e.g. backscattering and band-curvature effects, in the QSC case, and recent investigations have predicted super-Poissonian behavior in the noise of magnonic systems from dipole interactions.~\cite{kamraPRL16} These provide intriguing avenues to develop in future work. 

{\em Acknowledgments}:~The authors would like to thank A. Kamra and Y. Tserkovnyak for useful discussions. This research was supported by PSC-CUNY Research Award Program \#69366-00 47.


\begin{thebibliography}{54}%
\makeatletter
\providecommand \@ifxundefined [1]{%
 \@ifx{#1\undefined}
}%
\providecommand \@ifnum [1]{%
 \ifnum #1\expandafter \@firstoftwo
 \else \expandafter \@secondoftwo
 \fi
}%
\providecommand \@ifx [1]{%
 \ifx #1\expandafter \@firstoftwo
 \else \expandafter \@secondoftwo
 \fi
}%
\providecommand \natexlab [1]{#1}%
\providecommand \enquote  [1]{``#1''}%
\providecommand \bibnamefont  [1]{#1}%
\providecommand \bibfnamefont [1]{#1}%
\providecommand \citenamefont [1]{#1}%
\providecommand \href@noop [0]{\@secondoftwo}%
\providecommand \href [0]{\begingroup \@sanitize@url \@href}%
\providecommand \@href[1]{\@@startlink{#1}\@@href}%
\providecommand \@@href[1]{\endgroup#1\@@endlink}%
\providecommand \@sanitize@url [0]{\catcode `\\12\catcode `\$12\catcode
  `\&12\catcode `\#12\catcode `\^12\catcode `\_12\catcode `\%12\relax}%
\providecommand \@@startlink[1]{}%
\providecommand \@@endlink[0]{}%
\providecommand \url  [0]{\begingroup\@sanitize@url \@url }%
\providecommand \@url [1]{\endgroup\@href {#1}{\urlprefix }}%
\providecommand \urlprefix  [0]{URL }%
\providecommand \Eprint [0]{\href }%
\providecommand \doibase [0]{http://dx.doi.org/}%
\providecommand \selectlanguage [0]{\@gobble}%
\providecommand \bibinfo  [0]{\@secondoftwo}%
\providecommand \bibfield  [0]{\@secondoftwo}%
\providecommand \translation [1]{[#1]}%
\providecommand \BibitemOpen [0]{}%
\providecommand \bibitemStop [0]{}%
\providecommand \bibitemNoStop [0]{.\EOS\space}%
\providecommand \EOS [0]{\spacefactor3000\relax}%
\providecommand \BibitemShut  [1]{\csname bibitem#1\endcsname}%
\let\auto@bib@innerbib\@empty
\bibitem [{\citenamefont {Tserkovnyak}\ and\ \citenamefont
  {Brataas}(2001)}]{tserkovPRB01}%
  \BibitemOpen
  \bibfield  {author} {\bibinfo {author} {\bibfnamefont {Y.}~\bibnamefont
  {Tserkovnyak}}\ and\ \bibinfo {author} {\bibfnamefont {A.}~\bibnamefont
  {Brataas}},\ }\href@noop {} {\bibfield  {journal} {\bibinfo  {journal} {Phys.
  Rev. B}\ }\textbf {\bibinfo {volume} {64}},\ \bibinfo {eid} {214402}
  (\bibinfo {year} {2001})}\BibitemShut {NoStop}%
\bibitem [{\citenamefont {Mishchenko}\ and\ \citenamefont
  {Halperin}(2003)}]{mishchenkoPRB03}%
  \BibitemOpen
  \bibfield  {author} {\bibinfo {author} {\bibfnamefont {E.~G.}\ \bibnamefont
  {Mishchenko}}\ and\ \bibinfo {author} {\bibfnamefont {B.~I.}\ \bibnamefont
  {Halperin}},\ }\href@noop {} {\bibfield  {journal} {\bibinfo  {journal}
  {Phys. Rev. B}\ }\textbf {\bibinfo {volume} {68}},\ \bibinfo {eid} {045317}
  (\bibinfo {year} {2003})}\BibitemShut {NoStop}%
\bibitem [{\citenamefont {Lamacraft}(2004)}]{lamacraftPRB04}%
  \BibitemOpen
  \bibfield  {author} {\bibinfo {author} {\bibfnamefont {A.}~\bibnamefont
  {Lamacraft}},\ }\href {\doibase 10.1103/PhysRevB.69.081301} {\bibfield
  {journal} {\bibinfo  {journal} {Phys. Rev. B}\ }\textbf {\bibinfo {volume}
  {69}},\ \bibinfo {pages} {081301} (\bibinfo {year} {2004})}\BibitemShut
  {NoStop}%
\bibitem [{\citenamefont {Belzig}\ and\ \citenamefont
  {Zareyan}(2004)}]{belzigPRB04}%
  \BibitemOpen
  \bibfield  {author} {\bibinfo {author} {\bibfnamefont {W.}~\bibnamefont
  {Belzig}}\ and\ \bibinfo {author} {\bibfnamefont {M.}~\bibnamefont
  {Zareyan}},\ }\href@noop {} {\bibfield  {journal} {\bibinfo  {journal} {Phys.
  Rev. B}\ }\textbf {\bibinfo {volume} {69}},\ \bibinfo {pages} {140407}
  (\bibinfo {year} {2004})}\BibitemShut {NoStop}%
\bibitem [{\citenamefont {Foros}\ \emph {et~al.}(2005)\citenamefont {Foros},
  \citenamefont {Brataas}, \citenamefont {Tserkovnyak},\ and\ \citenamefont
  {Bauer}}]{forosPRL05}%
  \BibitemOpen
  \bibfield  {author} {\bibinfo {author} {\bibfnamefont {J.}~\bibnamefont
  {Foros}}, \bibinfo {author} {\bibfnamefont {A.}~\bibnamefont {Brataas}},
  \bibinfo {author} {\bibfnamefont {Y.}~\bibnamefont {Tserkovnyak}}, \ and\
  \bibinfo {author} {\bibfnamefont {G.~E.~W.}\ \bibnamefont {Bauer}},\
  }\href@noop {} {\bibfield  {journal} {\bibinfo  {journal} {Phys. Rev. Lett.}\
  }\textbf {\bibinfo {volume} {95}},\ \bibinfo {eid} {016601} (\bibinfo {year}
  {2005})}\BibitemShut {NoStop}%
\bibitem [{\citenamefont {Zareyan}\ and\ \citenamefont
  {Belzig}(2005)}]{zareyanEPL05}%
  \BibitemOpen
  \bibfield  {author} {\bibinfo {author} {\bibfnamefont {M.}~\bibnamefont
  {Zareyan}}\ and\ \bibinfo {author} {\bibfnamefont {W.}~\bibnamefont
  {Belzig}},\ }\href {http://stacks.iop.org/0295-5075/70/i=6/a=817} {\bibfield
  {journal} {\bibinfo  {journal} {EPL (Europhysics Letters)}\ }\textbf
  {\bibinfo {volume} {70}},\ \bibinfo {pages} {817} (\bibinfo {year}
  {2005})}\BibitemShut {NoStop}%
\bibitem [{\citenamefont {Chudnovskiy}\ \emph {et~al.}(2008)\citenamefont
  {Chudnovskiy}, \citenamefont {Swiebodzinski},\ and\ \citenamefont
  {Kamenev}}]{chudnovskiyPRL08}%
  \BibitemOpen
  \bibfield  {author} {\bibinfo {author} {\bibfnamefont {A.~L.}\ \bibnamefont
  {Chudnovskiy}}, \bibinfo {author} {\bibfnamefont {J.}~\bibnamefont
  {Swiebodzinski}}, \ and\ \bibinfo {author} {\bibfnamefont {A.}~\bibnamefont
  {Kamenev}},\ }\href@noop {} {\bibfield  {journal} {\bibinfo  {journal} {Phys.
  Rev. Lett.}\ }\textbf {\bibinfo {volume} {101}},\ \bibinfo {eid} {066601}
  (\bibinfo {year} {2008})}\BibitemShut {NoStop}%
\bibitem [{\citenamefont {Meair}\ \emph {et~al.}(2011)\citenamefont {Meair},
  \citenamefont {Stano},\ and\ \citenamefont {Jacquod}}]{meairPRB11}%
  \BibitemOpen
  \bibfield  {author} {\bibinfo {author} {\bibfnamefont {J.}~\bibnamefont
  {Meair}}, \bibinfo {author} {\bibfnamefont {P.}~\bibnamefont {Stano}}, \ and\
  \bibinfo {author} {\bibfnamefont {P.}~\bibnamefont {Jacquod}},\ }\href
  {\doibase 10.1103/PhysRevB.84.073302} {\bibfield  {journal} {\bibinfo
  {journal} {Phys. Rev. B}\ }\textbf {\bibinfo {volume} {84}},\ \bibinfo
  {pages} {073302} (\bibinfo {year} {2011})}\BibitemShut {NoStop}%
\bibitem [{\citenamefont {Arakawa}\ \emph {et~al.}(2015)\citenamefont
  {Arakawa}, \citenamefont {Shiogai}, \citenamefont {Ciorga}, \citenamefont
  {Utz}, \citenamefont {Schuh}, \citenamefont {Kohda}, \citenamefont {Nitta},
  \citenamefont {Bougeard}, \citenamefont {Weiss}, \citenamefont {Ono},\ and\
  \citenamefont {Kobayashi}}]{arakawaPRL15}%
  \BibitemOpen
  \bibfield  {author} {\bibinfo {author} {\bibfnamefont {T.}~\bibnamefont
  {Arakawa}}, \bibinfo {author} {\bibfnamefont {J.}~\bibnamefont {Shiogai}},
  \bibinfo {author} {\bibfnamefont {M.}~\bibnamefont {Ciorga}}, \bibinfo
  {author} {\bibfnamefont {M.}~\bibnamefont {Utz}}, \bibinfo {author}
  {\bibfnamefont {D.}~\bibnamefont {Schuh}}, \bibinfo {author} {\bibfnamefont
  {M.}~\bibnamefont {Kohda}}, \bibinfo {author} {\bibfnamefont
  {J.}~\bibnamefont {Nitta}}, \bibinfo {author} {\bibfnamefont
  {D.}~\bibnamefont {Bougeard}}, \bibinfo {author} {\bibfnamefont
  {D.}~\bibnamefont {Weiss}}, \bibinfo {author} {\bibfnamefont
  {T.}~\bibnamefont {Ono}}, \ and\ \bibinfo {author} {\bibfnamefont
  {K.}~\bibnamefont {Kobayashi}},\ }\href {\doibase
  10.1103/PhysRevLett.114.016601} {\bibfield  {journal} {\bibinfo  {journal}
  {Phys. Rev. Lett.}\ }\textbf {\bibinfo {volume} {114}},\ \bibinfo {pages}
  {016601} (\bibinfo {year} {2015})}\BibitemShut {NoStop}%
\bibitem [{\citenamefont {Wang}\ \emph {et~al.}(2004)\citenamefont {Wang},
  \citenamefont {Wang},\ and\ \citenamefont {Guo}}]{wangPRB04}%
  \BibitemOpen
  \bibfield  {author} {\bibinfo {author} {\bibfnamefont {B.}~\bibnamefont
  {Wang}}, \bibinfo {author} {\bibfnamefont {J.}~\bibnamefont {Wang}}, \ and\
  \bibinfo {author} {\bibfnamefont {H.}~\bibnamefont {Guo}},\ }\href {\doibase
  10.1103/PhysRevB.69.153301} {\bibfield  {journal} {\bibinfo  {journal} {Phys.
  Rev. B}\ }\textbf {\bibinfo {volume} {69}},\ \bibinfo {pages} {153301}
  (\bibinfo {year} {2004})}\BibitemShut {NoStop}%
\bibitem [{\citenamefont {Sauret}\ and\ \citenamefont
  {Feinberg}(2004)}]{sauretPRL04}%
  \BibitemOpen
  \bibfield  {author} {\bibinfo {author} {\bibfnamefont {O.}~\bibnamefont
  {Sauret}}\ and\ \bibinfo {author} {\bibfnamefont {D.}~\bibnamefont
  {Feinberg}},\ }\href {\doibase 10.1103/PhysRevLett.92.106601} {\bibfield
  {journal} {\bibinfo  {journal} {Phys. Rev. Lett.}\ }\textbf {\bibinfo
  {volume} {92}},\ \bibinfo {pages} {106601} (\bibinfo {year}
  {2004})}\BibitemShut {NoStop}%
\bibitem [{\citenamefont {Kamra}\ \emph {et~al.}(2014)\citenamefont {Kamra},
  \citenamefont {Witek}, \citenamefont {Meyer}, \citenamefont {Huebl},
  \citenamefont {Gepr\"ags}, \citenamefont {Gross}, \citenamefont {Bauer},\
  and\ \citenamefont {Goennenwein}}]{kamraPRB14}%
  \BibitemOpen
  \bibfield  {author} {\bibinfo {author} {\bibfnamefont {A.}~\bibnamefont
  {Kamra}}, \bibinfo {author} {\bibfnamefont {F.~P.}\ \bibnamefont {Witek}},
  \bibinfo {author} {\bibfnamefont {S.}~\bibnamefont {Meyer}}, \bibinfo
  {author} {\bibfnamefont {H.}~\bibnamefont {Huebl}}, \bibinfo {author}
  {\bibfnamefont {S.}~\bibnamefont {Gepr\"ags}}, \bibinfo {author}
  {\bibfnamefont {R.}~\bibnamefont {Gross}}, \bibinfo {author} {\bibfnamefont
  {G.~E.~W.}\ \bibnamefont {Bauer}}, \ and\ \bibinfo {author} {\bibfnamefont
  {S.~T.~B.}\ \bibnamefont {Goennenwein}},\ }\href {\doibase
  10.1103/PhysRevB.90.214419} {\bibfield  {journal} {\bibinfo  {journal} {Phys.
  Rev. B}\ }\textbf {\bibinfo {volume} {90}},\ \bibinfo {pages} {214419}
  (\bibinfo {year} {2014})}\BibitemShut {NoStop}%
\bibitem [{\citenamefont {Kamra}\ and\ \citenamefont
  {Belzig}(2016)}]{kamraPRL16}%
  \BibitemOpen
  \bibfield  {author} {\bibinfo {author} {\bibfnamefont {A.}~\bibnamefont
  {Kamra}}\ and\ \bibinfo {author} {\bibfnamefont {W.}~\bibnamefont {Belzig}},\
  }\href {\doibase 10.1103/PhysRevLett.116.146601} {\bibfield  {journal}
  {\bibinfo  {journal} {Phys. Rev. Lett.}\ }\textbf {\bibinfo {volume} {116}},\
  \bibinfo {pages} {146601} (\bibinfo {year} {2016})}\BibitemShut {NoStop}%
\bibitem [{\citenamefont {{Kamra}}\ and\ \citenamefont
  {{Belzig}}(2017)}]{kamraCM17}%
  \BibitemOpen
  \bibfield  {author} {\bibinfo {author} {\bibfnamefont {A.}~\bibnamefont
  {{Kamra}}}\ and\ \bibinfo {author} {\bibfnamefont {W.}~\bibnamefont
  {{Belzig}}},\ }\href@noop {} {} \Eprint
  {http://arxiv.org/abs/1706.07118} {arXiv:1706.07118}
  \BibitemShut {NoStop}%
\bibitem [{\citenamefont {{Matsuo}}\ \emph {et~al.}(2017)\citenamefont
  {{Matsuo}}, \citenamefont {{Ohnuma}}, \citenamefont {{Kato}},\ and\
  \citenamefont {{Maekawa}}}]{matsuoCM17}%
  \BibitemOpen
  \bibfield  {author} {\bibinfo {author} {\bibfnamefont {M.}~\bibnamefont
  {{Matsuo}}}, \bibinfo {author} {\bibfnamefont {Y.}~\bibnamefont {{Ohnuma}}},
  \bibinfo {author} {\bibfnamefont {T.}~\bibnamefont {{Kato}}}, \ and\ \bibinfo
  {author} {\bibfnamefont {S.}~\bibnamefont {{Maekawa}}},\ }\href@noop {}
 \Eprint {http://arxiv.org/abs/1711.00237} {arXiv:1711.00237
  } \BibitemShut {NoStop}%
\bibitem [{\citenamefont {Kajiwara}\ \emph {et~al.}(2010)\citenamefont
  {Kajiwara}, \citenamefont {Harii}, \citenamefont {Takahashi}, \citenamefont
  {Ohe}, \citenamefont {Uchida}, \citenamefont {Mizuguchi}, \citenamefont
  {Umezawa}, \citenamefont {Kawai}, \citenamefont {Ando}, \citenamefont
  {Takanashi}, \citenamefont {Maekawa},\ and\ \citenamefont
  {Saitoh}}]{kajiwaraNAT10}%
  \BibitemOpen
  \bibfield  {author} {\bibinfo {author} {\bibfnamefont {Y.}~\bibnamefont
  {Kajiwara}}, \bibinfo {author} {\bibfnamefont {K.}~\bibnamefont {Harii}},
  \bibinfo {author} {\bibfnamefont {S.}~\bibnamefont {Takahashi}}, \bibinfo
  {author} {\bibfnamefont {J.}~\bibnamefont {Ohe}}, \bibinfo {author}
  {\bibfnamefont {K.}~\bibnamefont {Uchida}}, \bibinfo {author} {\bibfnamefont
  {M.}~\bibnamefont {Mizuguchi}}, \bibinfo {author} {\bibfnamefont
  {H.}~\bibnamefont {Umezawa}}, \bibinfo {author} {\bibfnamefont
  {H.}~\bibnamefont {Kawai}}, \bibinfo {author} {\bibfnamefont
  {K.}~\bibnamefont {Ando}}, \bibinfo {author} {\bibfnamefont {K.}~\bibnamefont
  {Takanashi}}, \bibinfo {author} {\bibfnamefont {S.}~\bibnamefont {Maekawa}},
  \ and\ \bibinfo {author} {\bibfnamefont {E.}~\bibnamefont {Saitoh}},\
  }\href@noop {} {\bibfield  {journal} {\bibinfo  {journal} {Nature}\ }\textbf
  {\bibinfo {volume} {464}},\ \bibinfo {pages} {262} (\bibinfo {year}
  {2010})}\BibitemShut {NoStop}%
\bibitem [{\citenamefont {Cornelissen}\ \emph {et~al.}(2015)\citenamefont
  {Cornelissen}, \citenamefont {Liu}, \citenamefont {Duine}, \citenamefont
  {Youssef},\ and\ \citenamefont {van Wees}}]{cornelissenNATP15}%
  \BibitemOpen
  \bibfield  {author} {\bibinfo {author} {\bibfnamefont {L.~J.}\ \bibnamefont
  {Cornelissen}}, \bibinfo {author} {\bibfnamefont {J.}~\bibnamefont {Liu}},
  \bibinfo {author} {\bibfnamefont {R.~A.}\ \bibnamefont {Duine}}, \bibinfo
  {author} {\bibfnamefont {J.~B.}\ \bibnamefont {Youssef}}, \ and\ \bibinfo
  {author} {\bibfnamefont {B.~J.}\ \bibnamefont {van Wees}},\ }\href@noop {}
  {\bibfield  {journal} {\bibinfo  {journal} {Nature Phys.}\ }\textbf {\bibinfo
  {volume} {11}},\ \bibinfo {pages} {1022} (\bibinfo {year}
  {2015})}\BibitemShut {NoStop}%
\bibitem [{\citenamefont {Goennenwein}\ \emph {et~al.}(2015)\citenamefont
  {Goennenwein}, \citenamefont {Schlitz}, \citenamefont {Pernpeintner},
  \citenamefont {Ganzhorn}, \citenamefont {Althammer}, \citenamefont {Gross},\
  and\ \citenamefont {Huebl}}]{goennenweinAPL15}%
  \BibitemOpen
  \bibfield  {author} {\bibinfo {author} {\bibfnamefont {S.~T.~B.}\
  \bibnamefont {Goennenwein}}, \bibinfo {author} {\bibfnamefont
  {R.}~\bibnamefont {Schlitz}}, \bibinfo {author} {\bibfnamefont
  {M.}~\bibnamefont {Pernpeintner}}, \bibinfo {author} {\bibfnamefont
  {K.}~\bibnamefont {Ganzhorn}}, \bibinfo {author} {\bibfnamefont
  {M.}~\bibnamefont {Althammer}}, \bibinfo {author} {\bibfnamefont
  {R.}~\bibnamefont {Gross}}, \ and\ \bibinfo {author} {\bibfnamefont
  {H.}~\bibnamefont {Huebl}},\ }\href {\doibase 10.1063/1.4935074} {\bibfield
  {journal} {\bibinfo  {journal} {Appl. Phys. Lett.}\ }\textbf {\bibinfo
  {volume} {107}},\ \bibinfo {pages} {172405} (\bibinfo {year}
  {2015})}\BibitemShut {NoStop}%
\bibitem [{\citenamefont {Li}\ \emph {et~al.}(2016)\citenamefont {Li},
  \citenamefont {Xu}, \citenamefont {Aldosary}, \citenamefont {Tang},
  \citenamefont {Lin}, \citenamefont {Zhang}, \citenamefont {Lake},\ and\
  \citenamefont {Shi}}]{liNATC16}%
  \BibitemOpen
  \bibfield  {author} {\bibinfo {author} {\bibfnamefont {J.}~\bibnamefont
  {Li}}, \bibinfo {author} {\bibfnamefont {Y.}~\bibnamefont {Xu}}, \bibinfo
  {author} {\bibfnamefont {M.}~\bibnamefont {Aldosary}}, \bibinfo {author}
  {\bibfnamefont {C.}~\bibnamefont {Tang}}, \bibinfo {author} {\bibfnamefont
  {Z.}~\bibnamefont {Lin}}, \bibinfo {author} {\bibfnamefont {S.}~\bibnamefont
  {Zhang}}, \bibinfo {author} {\bibfnamefont {R.}~\bibnamefont {Lake}}, \ and\
  \bibinfo {author} {\bibfnamefont {J.}~\bibnamefont {Shi}},\ }\href
  {http://dx.doi.org/10.1038/ncomms10858} {\bibfield  {journal} {\bibinfo
  {journal} {Nat. Commun.}\ }\textbf {\bibinfo {volume} {7}}, \bibinfo {pages} {10858} (\bibinfo {year}
  {2016})}\BibitemShut {NoStop}%
\bibitem [{\citenamefont {Jordan}\ and\ \citenamefont
  {Wigner}(1928)}]{jordanZP28}%
  \BibitemOpen
  \bibfield  {author} {\bibinfo {author} {\bibfnamefont {P.}~\bibnamefont
  {Jordan}}\ and\ \bibinfo {author} {\bibfnamefont {E.}~\bibnamefont
  {Wigner}},\ }\href {\doibase 10.1007/BF01331938} {\bibfield  {journal}
  {\bibinfo  {journal} {Z. Phys.}\ }\textbf {\bibinfo
  {volume} {47}},\ \bibinfo {pages} {631} (\bibinfo {year} {1928})}\BibitemShut
  {NoStop}%
\bibitem [{\citenamefont {Kl{\"u}mper}(2004)}]{klumperBOOK04}%
  \BibitemOpen
  \bibfield  {author} {\bibinfo {author} {\bibfnamefont {A.}~\bibnamefont
  {Kl{\"u}mper}},\ }\enquote {\bibinfo {title} {Integrability of quantum
  chains: Theory and applications to the spin-1/2 xxz chain},}\ in\ \href@noop
  {} {\emph {\bibinfo {booktitle} {Quantum Magnetism}}},\ \bibinfo {editor}
  {edited by\ \bibinfo {editor} {\bibfnamefont {U.}~\bibnamefont
  {Schollw{\"o}ck}}, \bibinfo {editor} {\bibfnamefont {J.}~\bibnamefont
  {Richter}}, \bibinfo {editor} {\bibfnamefont {D.~J.~J.}\ \bibnamefont
  {Farnell}}, \ and\ \bibinfo {editor} {\bibfnamefont {R.~F.}\ \bibnamefont
  {Bishop}}}\ (\bibinfo  {publisher} {Springer Berlin Heidelberg},\ \bibinfo
  {address} {Berlin, Heidelberg},\ \bibinfo {year} {2004})\ pp.\ \bibinfo
  {pages} {349--379}\BibitemShut {NoStop}%
\bibitem [{\citenamefont {Mikeska}\ and\ \citenamefont
  {Kolezhuk}(2004)}]{mikeskaBOOK04}%
  \BibitemOpen
  \bibfield  {author} {\bibinfo {author} {\bibfnamefont {H.-J.}\ \bibnamefont
  {Mikeska}}\ and\ \bibinfo {author} {\bibfnamefont {A.~K.}\ \bibnamefont
  {Kolezhuk}},\ }\enquote {\bibinfo {title} {One-dimensional magnetism},}\ in\
  \href {\doibase 10.1007/BFb0119591} {\emph {\bibinfo {booktitle} {Quantum
  Magnetism}}},\ \bibinfo {editor} {edited by\ \bibinfo {editor} {\bibfnamefont
  {U.}~\bibnamefont {Schollw{\"o}ck}}, \bibinfo {editor} {\bibfnamefont
  {J.}~\bibnamefont {Richter}}, \bibinfo {editor} {\bibfnamefont {D.~J.~J.}\
  \bibnamefont {Farnell}}, \ and\ \bibinfo {editor} {\bibfnamefont {R.~F.}\
  \bibnamefont {Bishop}}}\ (\bibinfo  {publisher} {Springer Berlin
  Heidelberg},\ \bibinfo {address} {Berlin, Heidelberg},\ \bibinfo {year}
  {2004})\ pp.\ \bibinfo {pages} {1--83}\BibitemShut {NoStop}%
\bibitem [{\citenamefont {Hammar}\ \emph {et~al.}(1999)\citenamefont {Hammar},
  \citenamefont {Stone}, \citenamefont {Reich}, \citenamefont {Broholm},
  \citenamefont {Gibson}, \citenamefont {Turnbull}, \citenamefont {Landee},\
  and\ \citenamefont {Oshikawa}}]{hammarPRB99}%
  \BibitemOpen
  \bibfield  {author} {\bibinfo {author} {\bibfnamefont {P.~R.}\ \bibnamefont
  {Hammar}}, \bibinfo {author} {\bibfnamefont {M.~B.}\ \bibnamefont {Stone}},
  \bibinfo {author} {\bibfnamefont {D.~H.}\ \bibnamefont {Reich}}, \bibinfo
  {author} {\bibfnamefont {C.}~\bibnamefont {Broholm}}, \bibinfo {author}
  {\bibfnamefont {P.~J.}\ \bibnamefont {Gibson}}, \bibinfo {author}
  {\bibfnamefont {M.~M.}\ \bibnamefont {Turnbull}}, \bibinfo {author}
  {\bibfnamefont {C.~P.}\ \bibnamefont {Landee}}, \ and\ \bibinfo {author}
  {\bibfnamefont {M.}~\bibnamefont {Oshikawa}},\ }\href {\doibase
  10.1103/PhysRevB.59.1008} {\bibfield  {journal} {\bibinfo  {journal} {Phys.
  Rev. B}\ }\textbf {\bibinfo {volume} {59}},\ \bibinfo {pages} {1008}
  (\bibinfo {year} {1999})}\BibitemShut {NoStop}%
\bibitem [{\citenamefont {Stone}\ \emph {et~al.}(2003)\citenamefont {Stone},
  \citenamefont {Reich}, \citenamefont {Broholm}, \citenamefont {Lefmann},
  \citenamefont {Rischel}, \citenamefont {Landee},\ and\ \citenamefont
  {Turnbull}}]{stonePRL03}%
  \BibitemOpen
  \bibfield  {author} {\bibinfo {author} {\bibfnamefont {M.~B.}\ \bibnamefont
  {Stone}}, \bibinfo {author} {\bibfnamefont {D.~H.}\ \bibnamefont {Reich}},
  \bibinfo {author} {\bibfnamefont {C.}~\bibnamefont {Broholm}}, \bibinfo
  {author} {\bibfnamefont {K.}~\bibnamefont {Lefmann}}, \bibinfo {author}
  {\bibfnamefont {C.}~\bibnamefont {Rischel}}, \bibinfo {author} {\bibfnamefont
  {C.~P.}\ \bibnamefont {Landee}}, \ and\ \bibinfo {author} {\bibfnamefont
  {M.~M.}\ \bibnamefont {Turnbull}},\ }\href {\doibase
  10.1103/PhysRevLett.91.037205} {\bibfield  {journal} {\bibinfo  {journal}
  {Phys. Rev. Lett.}\ }\textbf {\bibinfo {volume} {91}},\ \bibinfo {pages}
  {037205} (\bibinfo {year} {2003})}\BibitemShut {NoStop}%
\bibitem [{\citenamefont {Lancaster}\ \emph {et~al.}(2006)\citenamefont
  {Lancaster}, \citenamefont {Blundell}, \citenamefont {Brooks}, \citenamefont
  {Baker}, \citenamefont {Pratt}, \citenamefont {Manson}, \citenamefont
  {Landee},\ and\ \citenamefont {Baines}}]{lancasterPRB06}%
  \BibitemOpen
  \bibfield  {author} {\bibinfo {author} {\bibfnamefont {T.}~\bibnamefont
  {Lancaster}}, \bibinfo {author} {\bibfnamefont {S.~J.}\ \bibnamefont
  {Blundell}}, \bibinfo {author} {\bibfnamefont {M.~L.}\ \bibnamefont
  {Brooks}}, \bibinfo {author} {\bibfnamefont {P.~J.}\ \bibnamefont {Baker}},
  \bibinfo {author} {\bibfnamefont {F.~L.}\ \bibnamefont {Pratt}}, \bibinfo
  {author} {\bibfnamefont {J.~L.}\ \bibnamefont {Manson}}, \bibinfo {author}
  {\bibfnamefont {C.~P.}\ \bibnamefont {Landee}}, \ and\ \bibinfo {author}
  {\bibfnamefont {C.}~\bibnamefont {Baines}},\ }\href {\doibase
  10.1103/PhysRevB.73.020410} {\bibfield  {journal} {\bibinfo  {journal} {Phys.
  Rev. B}\ }\textbf {\bibinfo {volume} {73}},\ \bibinfo {pages} {020410}
  (\bibinfo {year} {2006})}\BibitemShut {NoStop}%
\bibitem [{\citenamefont {K\"uhne}\ \emph {et~al.}(2009)\citenamefont
  {K\"uhne}, \citenamefont {Klauss}, \citenamefont {Grossjohann}, \citenamefont
  {Brenig}, \citenamefont {Litterst}, \citenamefont {Reyes}, \citenamefont
  {Kuhns}, \citenamefont {Turnbull},\ and\ \citenamefont
  {Landee}}]{kuehnePRB09}%
  \BibitemOpen
  \bibfield  {author} {\bibinfo {author} {\bibfnamefont {H.}~\bibnamefont
  {K\"uhne}}, \bibinfo {author} {\bibfnamefont {H.-H.}\ \bibnamefont {Klauss}},
  \bibinfo {author} {\bibfnamefont {S.}~\bibnamefont {Grossjohann}}, \bibinfo
  {author} {\bibfnamefont {W.}~\bibnamefont {Brenig}}, \bibinfo {author}
  {\bibfnamefont {F.~J.}\ \bibnamefont {Litterst}}, \bibinfo {author}
  {\bibfnamefont {A.~P.}\ \bibnamefont {Reyes}}, \bibinfo {author}
  {\bibfnamefont {P.~L.}\ \bibnamefont {Kuhns}}, \bibinfo {author}
  {\bibfnamefont {M.~M.}\ \bibnamefont {Turnbull}}, \ and\ \bibinfo {author}
  {\bibfnamefont {C.~P.}\ \bibnamefont {Landee}},\ }\href {\doibase
  10.1103/PhysRevB.80.045110} {\bibfield  {journal} {\bibinfo  {journal} {Phys.
  Rev. B}\ }\textbf {\bibinfo {volume} {80}},\ \bibinfo {pages} {045110}
  (\bibinfo {year} {2009})}\BibitemShut {NoStop}%
\bibitem [{\citenamefont {K\"uhne}\ \emph {et~al.}(2011)\citenamefont
  {K\"uhne}, \citenamefont {Zvyagin}, \citenamefont {G\"unther}, \citenamefont
  {Reyes}, \citenamefont {Kuhns}, \citenamefont {Turnbull}, \citenamefont
  {Landee},\ and\ \citenamefont {Klauss}}]{kuehnePRB11}%
  \BibitemOpen
  \bibfield  {author} {\bibinfo {author} {\bibfnamefont {H.}~\bibnamefont
  {K\"uhne}}, \bibinfo {author} {\bibfnamefont {A.~A.}\ \bibnamefont
  {Zvyagin}}, \bibinfo {author} {\bibfnamefont {M.}~\bibnamefont {G\"unther}},
  \bibinfo {author} {\bibfnamefont {A.~P.}\ \bibnamefont {Reyes}}, \bibinfo
  {author} {\bibfnamefont {P.~L.}\ \bibnamefont {Kuhns}}, \bibinfo {author}
  {\bibfnamefont {M.~M.}\ \bibnamefont {Turnbull}}, \bibinfo {author}
  {\bibfnamefont {C.~P.}\ \bibnamefont {Landee}}, \ and\ \bibinfo {author}
  {\bibfnamefont {H.-H.}\ \bibnamefont {Klauss}},\ }\href {\doibase
  10.1103/PhysRevB.83.100407} {\bibfield  {journal} {\bibinfo  {journal} {Phys.
  Rev. B}\ }\textbf {\bibinfo {volume} {83}},\ \bibinfo {pages} {100407}
  (\bibinfo {year} {2011})}\BibitemShut {NoStop}%
\bibitem [{\citenamefont {Klanj\v{s}ek}(2014)}]{klanjsekP14}%
  \BibitemOpen
  \bibfield  {author} {\bibinfo {author} {\bibfnamefont {M.}~\bibnamefont
  {Klanj\v{s}ek}},\ }\href {\doibase 10.1103/Physics.7.74} {\bibfield  {journal} {\bibinfo  {journal}
  {Physics}\ }\textbf {\bibinfo {volume} {7}},\ \bibinfo {pages} {74}
  (\bibinfo {year} {2014})}\BibitemShut {NoStop}%
\bibitem [{\citenamefont {Hirobe}\ \emph {et~al.}(2017)\citenamefont {Hirobe},
  \citenamefont {Sato}, \citenamefont {Kawamata}, \citenamefont {Shiomi},
  \citenamefont {Uchida}, \citenamefont {Iguchi}, \citenamefont {Koike},
  \citenamefont {Maekawa},\ and\ \citenamefont {Saitoh}}]{hirobeNATP17}%
  \BibitemOpen
  \bibfield  {author} {\bibinfo {author} {\bibfnamefont {D.}~\bibnamefont
  {Hirobe}}, \bibinfo {author} {\bibfnamefont {M.}~\bibnamefont {Sato}},
  \bibinfo {author} {\bibfnamefont {T.}~\bibnamefont {Kawamata}}, \bibinfo
  {author} {\bibfnamefont {Y.}~\bibnamefont {Shiomi}}, \bibinfo {author}
  {\bibfnamefont {K.-i.}\ \bibnamefont {Uchida}}, \bibinfo {author}
  {\bibfnamefont {R.}~\bibnamefont {Iguchi}}, \bibinfo {author} {\bibfnamefont
  {Y.}~\bibnamefont {Koike}}, \bibinfo {author} {\bibfnamefont
  {S.}~\bibnamefont {Maekawa}}, \ and\ \bibinfo {author} {\bibfnamefont
  {E.}~\bibnamefont {Saitoh}},\ }\href@noop {} {\bibfield  {journal} {\bibinfo
  {journal} {Nature Phys.}\ }\textbf {\bibinfo {volume} {13}},\ \bibinfo
  {pages} {30} (\bibinfo {year} {2017})}\BibitemShut {NoStop}%
\bibitem [{\citenamefont {Heiblum}(2006)}]{heiblumPSS06}%
  \BibitemOpen
  \bibfield  {author} {\bibinfo {author} {\bibfnamefont {M.}~\bibnamefont
  {Heiblum}},\ }\href {\doibase 10.1002/pssb.200642237} {\bibfield  {journal}
  {\bibinfo  {journal} {Phys. Status Solidi B}\ }\textbf {\bibinfo {volume}
  {243}},\ \bibinfo {pages} {3604} (\bibinfo {year} {2006})}\BibitemShut
  {NoStop}%
\bibitem [{\citenamefont {Kane}\ and\ \citenamefont
  {Fisher}(2007)}]{kaneBOOK07}%
  \BibitemOpen
  \bibfield  {author} {\bibinfo {author} {\bibfnamefont {C.~L.}\ \bibnamefont
  {Kane}}\ and\ \bibinfo {author} {\bibfnamefont {M.~P.~A.}\ \bibnamefont
  {Fisher}},\ }\enquote {\bibinfo {title} {Edge-state transport},}\ in\ \href
  {\doibase 10.1002/9783527617258.ch4} {\emph {\bibinfo {booktitle}
  {Perspectives in Quantum Hall Effects}}}\ (\bibinfo  {publisher} {Wiley-VCH
  Verlag GmbH, Weinheim, Germany},\ \bibinfo {year} {2007})\ pp.\ \bibinfo {pages}
  {109--159}\BibitemShut {NoStop}%
\bibitem [{\citenamefont {Maekawa}\ \emph {et~al.}(2012)\citenamefont
  {Maekawa}, \citenamefont {Valenzuela}, \citenamefont {Saitoh},\ and\
  \citenamefont {Kimura}}]{maekawaBOOK12}%
  \BibitemOpen
  \bibfield  {author} {\bibinfo {author} {\bibfnamefont {S.}~\bibnamefont
  {Maekawa}}, \bibinfo {author} {\bibfnamefont {S.}~\bibnamefont {Valenzuela}},
  \bibinfo {author} {\bibfnamefont {E.}~\bibnamefont {Saitoh}}, \ and\ \bibinfo
  {author} {\bibfnamefont {T.}~\bibnamefont {Kimura}},\ }\href
  {https://books.google.com/books?id=--u7GSdsazoC} {\emph {\bibinfo {title}
  {Spin Current}}},\ Series on Semiconductor Science and Technology\ (\bibinfo
  {publisher} {OUP Oxford},\ \bibinfo {year} {2012})\BibitemShut {NoStop}%
\bibitem [{\citenamefont {Sinova}\ \emph {et~al.}(2015)\citenamefont {Sinova},
  \citenamefont {Valenzuela}, \citenamefont {Wunderlich}, \citenamefont
  {Back},\ and\ \citenamefont {Jungwirth}}]{sinovaRMP15}%
  \BibitemOpen
  \bibfield  {author} {\bibinfo {author} {\bibfnamefont {J.}~\bibnamefont
  {Sinova}}, \bibinfo {author} {\bibfnamefont {S.~O.}\ \bibnamefont
  {Valenzuela}}, \bibinfo {author} {\bibfnamefont {J.}~\bibnamefont
  {Wunderlich}}, \bibinfo {author} {\bibfnamefont {C.~H.}\ \bibnamefont
  {Back}}, \ and\ \bibinfo {author} {\bibfnamefont {T.}~\bibnamefont
  {Jungwirth}},\ }\href {\doibase 10.1103/RevModPhys.87.1213} {\bibfield
  {journal} {\bibinfo  {journal} {Rev. Mod. Phys.}\ }\textbf {\bibinfo {volume}
  {87}},\ \bibinfo {pages} {1213} (\bibinfo {year} {2015})}\BibitemShut
  {NoStop}%
\bibitem [{\citenamefont {Brataas}\ \emph {et~al.}(2012)\citenamefont
  {Brataas}, \citenamefont {Kent},\ and\ \citenamefont {Ohno}}]{brataasNATM12}%
  \BibitemOpen
  \bibfield  {author} {\bibinfo {author} {\bibfnamefont {A.}~\bibnamefont
  {Brataas}}, \bibinfo {author} {\bibfnamefont {A.~D.}\ \bibnamefont {Kent}}, \
  and\ \bibinfo {author} {\bibfnamefont {H.}~\bibnamefont {Ohno}},\ }\href@noop
  {} {\bibfield  {journal} {\bibinfo  {journal} {Nat Mater}\ }\textbf {\bibinfo
  {volume} {11}},\ \bibinfo {pages} {372} (\bibinfo {year} {2012})}\BibitemShut
  {NoStop}%
\bibitem [{\citenamefont {Haney}\ \emph {et~al.}(2013)\citenamefont {Haney},
  \citenamefont {Lee}, \citenamefont {Lee}, \citenamefont {Manchon},\ and\
  \citenamefont {Stiles}}]{haneyPRB13}%
  \BibitemOpen
  \bibfield  {author} {\bibinfo {author} {\bibfnamefont {P.~M.}\ \bibnamefont
  {Haney}}, \bibinfo {author} {\bibfnamefont {H.-W.}\ \bibnamefont {Lee}},
  \bibinfo {author} {\bibfnamefont {K.-J.}\ \bibnamefont {Lee}}, \bibinfo
  {author} {\bibfnamefont {A.}~\bibnamefont {Manchon}}, \ and\ \bibinfo
  {author} {\bibfnamefont {M.~D.}\ \bibnamefont {Stiles}},\ }\href {\doibase
  10.1103/PhysRevB.87.174411} {\bibfield  {journal} {\bibinfo  {journal} {Phys.
  Rev. B}\ }\textbf {\bibinfo {volume} {87}},\ \bibinfo {pages} {174411}
  (\bibinfo {year} {2013})}\BibitemShut {NoStop}%
\bibitem [{\citenamefont {Tserkovnyak}\ and\ \citenamefont
  {Bender}(2014)}]{tserkovnyakPRB14}%
  \BibitemOpen
  \bibfield  {author} {\bibinfo {author} {\bibfnamefont {Y.}~\bibnamefont
  {Tserkovnyak}}\ and\ \bibinfo {author} {\bibfnamefont {S.~A.}\ \bibnamefont
  {Bender}},\ }\href {\doibase 10.1103/PhysRevB.90.014428} {\bibfield
  {journal} {\bibinfo  {journal} {Phys. Rev. B}\ }\textbf {\bibinfo {volume}
  {90}},\ \bibinfo {pages} {014428} (\bibinfo {year} {2014})}\BibitemShut
  {NoStop}%
\bibitem [{\citenamefont {Gogolin}(1998)}]{gogolinBOOK98}%
  \BibitemOpen
  \bibfield  {author} {\bibinfo {author} {\bibfnamefont {A.~O.}\ \bibnamefont
  {Gogolin}},\ }\href@noop {} {\emph {\bibinfo {title} {Bosonization and
  Strongly Correlated Systems}}}\ (\bibinfo  {publisher} {Cambridge University
  Press},\ \bibinfo {year} {1998})\BibitemShut {NoStop}%
\bibitem [{\citenamefont {Miranda}(2003)}]{mirandaBJP03}%
  \BibitemOpen
  \bibfield  {author} {\bibinfo {author} {\bibfnamefont {E.}~\bibnamefont
  {Miranda}},\ }\href@noop {} {\bibfield  {journal} {\bibinfo  {journal}
  {{Brazilian Journal of Physics}}\ }\textbf {\bibinfo {volume} {33}},\
  \bibinfo {pages} {3 } (\bibinfo {year} {2003})}\BibitemShut {NoStop}%
\bibitem [{\citenamefont {Giamarchi}(2004)}]{giamarchiBOOK04}%
  \BibitemOpen
  \bibfield  {author} {\bibinfo {author} {\bibfnamefont {T.}~\bibnamefont
  {Giamarchi}},\ }\href@noop {} {\emph {\bibinfo {title} {Quantum Physics in
  One Dimension}}}\ (\bibinfo  {publisher} {Oxford University Press},\ \bibinfo
  {address} {Oxford},\ \bibinfo {year} {2004})\BibitemShut {NoStop}%
\bibitem [{\citenamefont {Johnson}\ \emph {et~al.}(1973)\citenamefont
  {Johnson}, \citenamefont {Krinsky},\ and\ \citenamefont
  {McCoy}}]{johnsonPRA73}%
  \BibitemOpen
  \bibfield  {author} {\bibinfo {author} {\bibfnamefont {J.~D.}\ \bibnamefont
  {Johnson}}, \bibinfo {author} {\bibfnamefont {S.}~\bibnamefont {Krinsky}}, \
  and\ \bibinfo {author} {\bibfnamefont {B.~M.}\ \bibnamefont {McCoy}},\ }\href
  {\doibase 10.1103/PhysRevA.8.2526} {\bibfield  {journal} {\bibinfo  {journal}
  {Phys. Rev. A}\ }\textbf {\bibinfo {volume} {8}},\ \bibinfo {pages} {2526}
  (\bibinfo {year} {1973})}\BibitemShut {NoStop}%
\bibitem [{Note1()}]{Note1}%
  \BibitemOpen
  \bibinfo {note} {For a full explication of a finite edge in a Luttinger
  liquid, see section $10.1$ of Ref.~\protect \rev@citealpnum
  {giamarchiBOOK04}.}\BibitemShut {Stop}%
\bibitem [{\citenamefont {Rammer}(2007)}]{rammerBOOK07}%
  \BibitemOpen
  \bibfield  {author} {\bibinfo {author} {\bibfnamefont {J.}~\bibnamefont
  {Rammer}},\ }\href {https://books.google.com/books?id=A7TbrAm5Wq0C} {\emph
  {\bibinfo {title} {Quantum Field Theory of Non-equilibrium States}}}\
  (\bibinfo  {publisher} {Cambridge University Press},\ \bibinfo {year}
  {2007})\BibitemShut {NoStop}%
\bibitem [{\citenamefont {Kamenev}(2011)}]{kamenevBOOK11}%
  \BibitemOpen
  \bibfield  {author} {\bibinfo {author} {\bibfnamefont {A.}~\bibnamefont
  {Kamenev}},\ }\href@noop {} {\emph {\bibinfo {title} {Field Theory of
  Non-Equilibrium Systems}}}\ (\bibinfo  {publisher} {Cambridge University
  Press},\ \bibinfo {address} {Cambridge},\ \bibinfo {year} {2011})\BibitemShut
  {NoStop}%
\bibitem [{\citenamefont {Cornelissen}\ \emph {et~al.}(2016)\citenamefont
  {Cornelissen}, \citenamefont {Peters}, \citenamefont {Bauer}, \citenamefont
  {Duine},\ and\ \citenamefont {van Wees}}]{cornelissenPRB16}%
  \BibitemOpen
  \bibfield  {author} {\bibinfo {author} {\bibfnamefont {L.~J.}\ \bibnamefont
  {Cornelissen}}, \bibinfo {author} {\bibfnamefont {K.~J.~H.}\ \bibnamefont
  {Peters}}, \bibinfo {author} {\bibfnamefont {G.~E.~W.}\ \bibnamefont
  {Bauer}}, \bibinfo {author} {\bibfnamefont {R.~A.}\ \bibnamefont {Duine}}, \
  and\ \bibinfo {author} {\bibfnamefont {B.~J.}\ \bibnamefont {van Wees}},\
  }\href {\doibase 10.1103/PhysRevB.94.014412} {\bibfield  {journal} {\bibinfo
  {journal} {Phys. Rev. B}\ }\textbf {\bibinfo {volume} {94}},\ \bibinfo
  {pages} {014412} (\bibinfo {year} {2016})}\BibitemShut {NoStop}%
\bibitem [{\citenamefont {Shan}\ \emph {et~al.}(2016)\citenamefont {Shan},
  \citenamefont {Cornelissen}, \citenamefont {Vlietstra}, \citenamefont
  {Ben~Youssef}, \citenamefont {Kuschel}, \citenamefont {Duine},\ and\
  \citenamefont {van Wees}}]{shanPRB16}%
  \BibitemOpen
  \bibfield  {author} {\bibinfo {author} {\bibfnamefont {J.}~\bibnamefont
  {Shan}}, \bibinfo {author} {\bibfnamefont {L.~J.}\ \bibnamefont
  {Cornelissen}}, \bibinfo {author} {\bibfnamefont {N.}~\bibnamefont
  {Vlietstra}}, \bibinfo {author} {\bibfnamefont {J.}~\bibnamefont
  {Ben~Youssef}}, \bibinfo {author} {\bibfnamefont {T.}~\bibnamefont
  {Kuschel}}, \bibinfo {author} {\bibfnamefont {R.~A.}\ \bibnamefont {Duine}},
  \ and\ \bibinfo {author} {\bibfnamefont {B.~J.}\ \bibnamefont {van Wees}},\
  }\href {\doibase 10.1103/PhysRevB.94.174437} {\bibfield  {journal} {\bibinfo
  {journal} {Phys. Rev. B}\ }\textbf {\bibinfo {volume} {94}},\ \bibinfo
  {pages} {174437} (\bibinfo {year} {2016})}\BibitemShut {NoStop}%
\bibitem [{\citenamefont {R\"uckriegel}\ \emph {et~al.}(2014)\citenamefont
  {R\"uckriegel}, \citenamefont {Kopietz}, \citenamefont {Bozhko},
  \citenamefont {Serga},\ and\ \citenamefont {Hillebrands}}]{rueckriegelPRB14}%
  \BibitemOpen
  \bibfield  {author} {\bibinfo {author} {\bibfnamefont {A.}~\bibnamefont
  {R\"uckriegel}}, \bibinfo {author} {\bibfnamefont {P.}~\bibnamefont
  {Kopietz}}, \bibinfo {author} {\bibfnamefont {D.~A.}\ \bibnamefont {Bozhko}},
  \bibinfo {author} {\bibfnamefont {A.~A.}\ \bibnamefont {Serga}}, \ and\
  \bibinfo {author} {\bibfnamefont {B.}~\bibnamefont {Hillebrands}},\ }\href
  {\doibase 10.1103/PhysRevB.89.184413} {\bibfield  {journal} {\bibinfo
  {journal} {Phys. Rev. B}\ }\textbf {\bibinfo {volume} {89}},\ \bibinfo
  {pages} {184413} (\bibinfo {year} {2014})}\BibitemShut {NoStop}%
\bibitem [{\citenamefont {Holstein}\ and\ \citenamefont
  {Primakoff}(1940)}]{holsteinPR40}%
  \BibitemOpen
  \bibfield  {author} {\bibinfo {author} {\bibfnamefont {T.}~\bibnamefont
  {Holstein}}\ and\ \bibinfo {author} {\bibfnamefont {H.}~\bibnamefont
  {Primakoff}},\ }\href@noop {} {\bibfield  {journal} {\bibinfo  {journal}
  {Phys. Rev.}\ }\textbf {\bibinfo {volume} {58}},\ \bibinfo {pages} {1098}
  (\bibinfo {year} {1940})}\BibitemShut {NoStop}%
\bibitem [{Note2()}]{Note2}%
  \BibitemOpen
  \bibinfo {note} {Ref.~\protect \rev@citealpnum {kamraPRL16} has calculated
  shot noise in the spin current generated via spin pumping with a mono-domain
  ferromagnet and reported super-Poissonian shot noise resulting from dipolar
  interactions.}\BibitemShut {Stop}%
\bibitem [{\citenamefont {de~C.~Chamon}\ and\ \citenamefont
  {Wen}(1993)}]{chamonPRL93}%
  \BibitemOpen
  \bibfield  {author} {\bibinfo {author} {\bibfnamefont {C.}~\bibnamefont
  {de~C.~Chamon}}\ and\ \bibinfo {author} {\bibfnamefont {X.~G.}\ \bibnamefont
  {Wen}},\ }\href {\doibase 10.1103/PhysRevLett.70.2605} {\bibfield  {journal}
  {\bibinfo  {journal} {Phys. Rev. Lett.}\ }\textbf {\bibinfo {volume} {70}},\
  \bibinfo {pages} {2605} (\bibinfo {year} {1993})}\BibitemShut {NoStop}%
\bibitem [{\citenamefont {Kane}\ and\ \citenamefont
  {Fisher}(1994)}]{kanePRL94}%
  \BibitemOpen
  \bibfield  {author} {\bibinfo {author} {\bibfnamefont {C.~L.}\ \bibnamefont
  {Kane}}\ and\ \bibinfo {author} {\bibfnamefont {M.~P.~A.}\ \bibnamefont
  {Fisher}},\ }\href {\doibase 10.1103/PhysRevLett.72.724} {\bibfield
  {journal} {\bibinfo  {journal} {Phys. Rev. Lett.}\ }\textbf {\bibinfo
  {volume} {72}},\ \bibinfo {pages} {724} (\bibinfo {year} {1994})}\BibitemShut
  {NoStop}%
\bibitem [{\citenamefont {Ono}\ \emph {et~al.}(2002)\citenamefont {Ono},
  \citenamefont {Austing}, \citenamefont {Tokura},\ and\ \citenamefont
  {Tarucha}}]{onoSCI02}%
  \BibitemOpen
  \bibfield  {author} {\bibinfo {author} {\bibfnamefont {K.}~\bibnamefont
  {Ono}}, \bibinfo {author} {\bibfnamefont {D.~G.}\ \bibnamefont {Austing}},
  \bibinfo {author} {\bibfnamefont {Y.}~\bibnamefont {Tokura}}, \ and\ \bibinfo
  {author} {\bibfnamefont {S.}~\bibnamefont {Tarucha}},\ }\href@noop {}
  {\bibfield  {journal} {\bibinfo  {journal} {Science}\ }\textbf {\bibinfo
  {volume} {297}},\ \bibinfo {pages} {1313} (\bibinfo {year}
  {2002})}\BibitemShut {NoStop}%
\bibitem [{\citenamefont {Takei}\ \emph {et~al.}(2015)\citenamefont {Takei},
  \citenamefont {Rosenow},\ and\ \citenamefont {Stern}}]{takeiPRB15}%
  \BibitemOpen
  \bibfield  {author} {\bibinfo {author} {\bibfnamefont {S.}~\bibnamefont
  {Takei}}, \bibinfo {author} {\bibfnamefont {B.}~\bibnamefont {Rosenow}}, \
  and\ \bibinfo {author} {\bibfnamefont {A.}~\bibnamefont {Stern}},\ }\href
  {\doibase 10.1103/PhysRevB.91.241104} {\bibfield  {journal} {\bibinfo
  {journal} {Phys. Rev. B}\ }\textbf {\bibinfo {volume} {91}},\ \bibinfo
  {pages} {241104} (\bibinfo {year} {2015})}\BibitemShut {NoStop}%
\bibitem [{\citenamefont {Mosendz}\ \emph {et~al.}(2010)\citenamefont
  {Mosendz}, \citenamefont {Vlaminck}, \citenamefont {Pearson}, \citenamefont
  {Fradin}, \citenamefont {Bauer}, \citenamefont {Bader},\ and\ \citenamefont
  {Hoffmann}}]{mosendzPRB10}%
  \BibitemOpen
  \bibfield  {author} {\bibinfo {author} {\bibfnamefont {O.}~\bibnamefont
  {Mosendz}}, \bibinfo {author} {\bibfnamefont {V.}~\bibnamefont {Vlaminck}},
  \bibinfo {author} {\bibfnamefont {J.~E.}\ \bibnamefont {Pearson}}, \bibinfo
  {author} {\bibfnamefont {F.~Y.}\ \bibnamefont {Fradin}}, \bibinfo {author}
  {\bibfnamefont {G.~E.~W.}\ \bibnamefont {Bauer}}, \bibinfo {author}
  {\bibfnamefont {S.~D.}\ \bibnamefont {Bader}}, \ and\ \bibinfo {author}
  {\bibfnamefont {A.}~\bibnamefont {Hoffmann}},\ }\href {\doibase
  10.1103/PhysRevB.82.214403} {\bibfield  {journal} {\bibinfo  {journal} {Phys.
  Rev. B}\ }\textbf {\bibinfo {volume} {82}},\ \bibinfo {pages} {214403}
  (\bibinfo {year} {2010})}\BibitemShut {NoStop}%
\bibitem [{\citenamefont {Johnston}(1997)}]{johnstonBOOK97}%
  \BibitemOpen
  \bibfield  {author} {\bibinfo {author} {\bibfnamefont {D.~C.}\ \bibnamefont
  {Johnston}},\ }\href {\doibase https://doi.org/10.1016/S1567-2719(97)10005-1}
  {\emph {\bibinfo {title} {Chapter 1 Normal-state magnetic properties of
  single-layer cuprate high-temperature superconductors and related
  materials}}},\ \bibinfo {series} {Handbook of Magnetic Materials},
  Vol.~\bibinfo {volume} {10}\ (\bibinfo  {publisher} {Elsevier, New York},\ \bibinfo
  {year} {1997})\ pp.\ \bibinfo {pages} {1 -- 237}\BibitemShut {NoStop}%
\end{thebibliography}
\end{document}